%% file: main.tex
\documentclass[12pt]{article}

\usepackage[utf8]{inputenc}
\usepackage{amsmath, amssymb, amsthm, mathtools, bbm, graphicx, fullpage, caption, subcaption, multirow, url, authblk, hyperref}
\usepackage[dvipsnames]{xcolor}
\usepackage[normalem]{ulem}
\usepackage{rotating}
\usepackage{setspace}
\usepackage{booktabs}
\input{preamble.tex}

\title{From PhysioNet to Foundation Models - A history and potential futures}

\author[1,2]{\small{Gari D. Clifford}}
\affil[1]{\small{Department of Biomedical Informatics, Emory University}}
\affil[2]{\small{Department of Biomedical Engineering, Georgia Institute of Technology \& Emory University}}
\affil[3]{\small{PhysioNet, The Research Resource for Complex Physiologic Signals, Harvard University}}

\date{1 January 2026}

\begin{document}

\maketitle

\blkfootnote{
The work described in this publication was supported in part by the National Institute of Biomedical Imaging and Bioengineering (NIBIB) and by the Director’s Office of Data Science Strategy (ODSS) of the National Institutes of Health (NIH) under Award Number R01EB030362. 
Funding for the  George B. Moody PhysioNet Challenges has also been provided by Alivecor Inc., Amazon Web Services, Google, the Gordon and Betty Moore Foundation, the IEEE Signal Processing Society, and Mathworks Inc. In addition, the Federation of American Societies for Experimental Biology (FASEB) specifically recognized the Challenges as a recipient of the ``Distinguished Achievement Award for Data Reuse'', as part of the inaugural DataWorks! prize in 2022. The prize money has also been used to support the Challenges. %This award acknowledges the Challenges' long history and success in encouraging the reuse of data to improve health and healthcare.
The content is solely the responsibility of the author and does not necessarily represent the official views of the NIH, my current and past employers, colleagues, or sponsors. 
The author declares no conflict of interest.

For information regarding this article, please contact the author via email. Address: Department of Biomedical Informatics, Emory University, Woodruff Memorial Research Building, 101 Woodruff Circle, 4th Floor East, Atlanta, GA 30322, USA. Phone: (404)-727-4631. Email: gari@physionet.org
}

%%%%%%%%%%%%%%%%%%%%%%%%%%%%%%%%%%%%%%%%
%
% Abstract
%
%%%%%%%%%%%%%%%%%%%%%%%%%%%%%%%%%%%%%%%%

\newpage

\begin{abstract}

  Over the last 35 years, the sharing of medical data and models for research has evolved from sneakernet to the internet - from mailing magnetic tapes and compact discs of a handful of well-curated recordings, to the high-speed download of relatively comprehensive 
  %(almost {\it ad hoc} organized) 
  hospital databases. 
  More recently, the fervor around the potential for modern machine learning and `AI' to catapult us into the next industrial revolution has led to a seemingly insatiable desire to pump almost any source of data into large models. Although this has great potential, it also presents a whole set of new challenges.

In this article I examine these trends over the last 30 years, drawing on examples from the world of physiology, and in particular, cardiology, since it is one of the oldest data-intensive fields, and is undergoing a renaissance in the context of machine learning. From the early days of computerized cardiology, the Research Resource for Complex Physiologic Signals ({\it PhysioNet}) has been at the cutting edge of this field. 
This article, therefore, includes much of the Resource's history and the contributions of many key people involved, drawn from almost three decades of firsthand experience of co-developing elements of the Resource with its founders. 

I address what I perceive to be the most promising future directions for the PhysioNet Resource, and more generally, the growing issues and opportunities around dissemination and use of massive physiological databases, associated open access code, and public competitions, together with potential solutions to the key issues that face our field. 
Topics range from how we should approach foundation models in the context of the rapidly growing AI carbon footprint, to the potential of {\it Tiny-ML} and edge computing.
I also cover issues around prizes and incentives, funding models, and scientific repeatability, 
as well as how we might address these issues by leveraging the PhysioNet Challenges, consistent with the philosophy of open-access from the early days of the PhysioNet Resource. 


\end{abstract}

%%%%%%%%%%%%%%%%%%%%%%%%%%%%%%%%%%%%%%%%%%%%%%%%%
\section{Introduction}
\input{0_introduction}
%%%%%%%%%%%%%%%%%%%%%%%%%%%%%%%%%%%%%%%%%%

\section{Some history on open access science}\label{sec:history}
\input{1_history}

\section{Tutorials, publishing, trust, and misaligned incentives}\label{sec:discussion}
\input{4_discussion}
%%%%%%%%%%%%%%%%%%%%%%%%%%%%%%%%%%%%%%%%%%

\section{Recent Developments using `AI'}\label{section:RecentAI}

\input{2_methods}
%%%%%%%%%%%%%%%%%%%%%%%%%%%%%%%%%%%%%%%%%%

%\section{Tiny ML and Edge AI}\label{sec:results}
%\input{3_results}
%%%%%%%%%%%%%%%%%%%%%%%%%%%%%%%%%%%%%%%%%%

\section{The Future? Large Models, Public Data and the role of PhysioNet and public other resources}

\input{5_conclusion}
%%%%%%%%%%%%%%%%%%%%%%%%%%%%%%%%%%%%%%%%%%

\section{Final Thoughts: TLDR}
\input{6_TLDR}

\section*{Appendix}
\input{7_Appendix}

\bibliographystyle{unsrt}
\bibliography{refs,challengerefs,competitions,physionet}
\end{document}

%% file: preamble.tex
\newcommand{\blkfootnote}[1]{\begingroup\renewcommand\thefootnote{}\footnote{#1}\addtocounter{footnote}{-1}\endgroup}

%% file: 0_introduction.tex
We are at a critical point for many industries, particularly in medicine. Artificial Intelligence (AI) is being hyped as a transformative technology across many fields, and it is likely that we will look back on this period as an inflection point. Of course, anyone who tries to predict the future of society and technological development is also likely to be wrong, but the rate of change in automated decision support in medicine feels qualitatively different today, and there is good reason to be optimistic. The medical field has often been criticized for its resistance to change. For example, many subspecialties operate within paradigms shaped by thoughtful expert consensus that, while valuable, may inadvertently limit adaptation to new evidence or technologies.
Data-driven approaches to optimizing clinical decision-making offer much promise for advancing the field.  However, the speed of change is a cause for concern. To invert the old Facebook motto, when things move fast, things break. So we need to think carefully about how we approach this new paradigm and set up some guide rails.  

This article started out as an invited talk at Computing in Cardiology 2024 \cite{GariDClifford_2024_PastPresent}, and I was then asked to turn this into a longer journal article. Since then, it has morphed into both a mini history of PhysioNet, and a landscaping of where I think we should be going next with big data, foundation models and open science. While my experience is confined to the biomedical domain, I believe a lot of what I have written could be more broadly applicable to the general AI gold rush. In particular, I have drawn on my experience of supporting PhysioNet for the last quarter of a century, and my work in global health over the same period.

\begin{figure}[h!]
    \centering
  \includegraphics[scale=0.12]{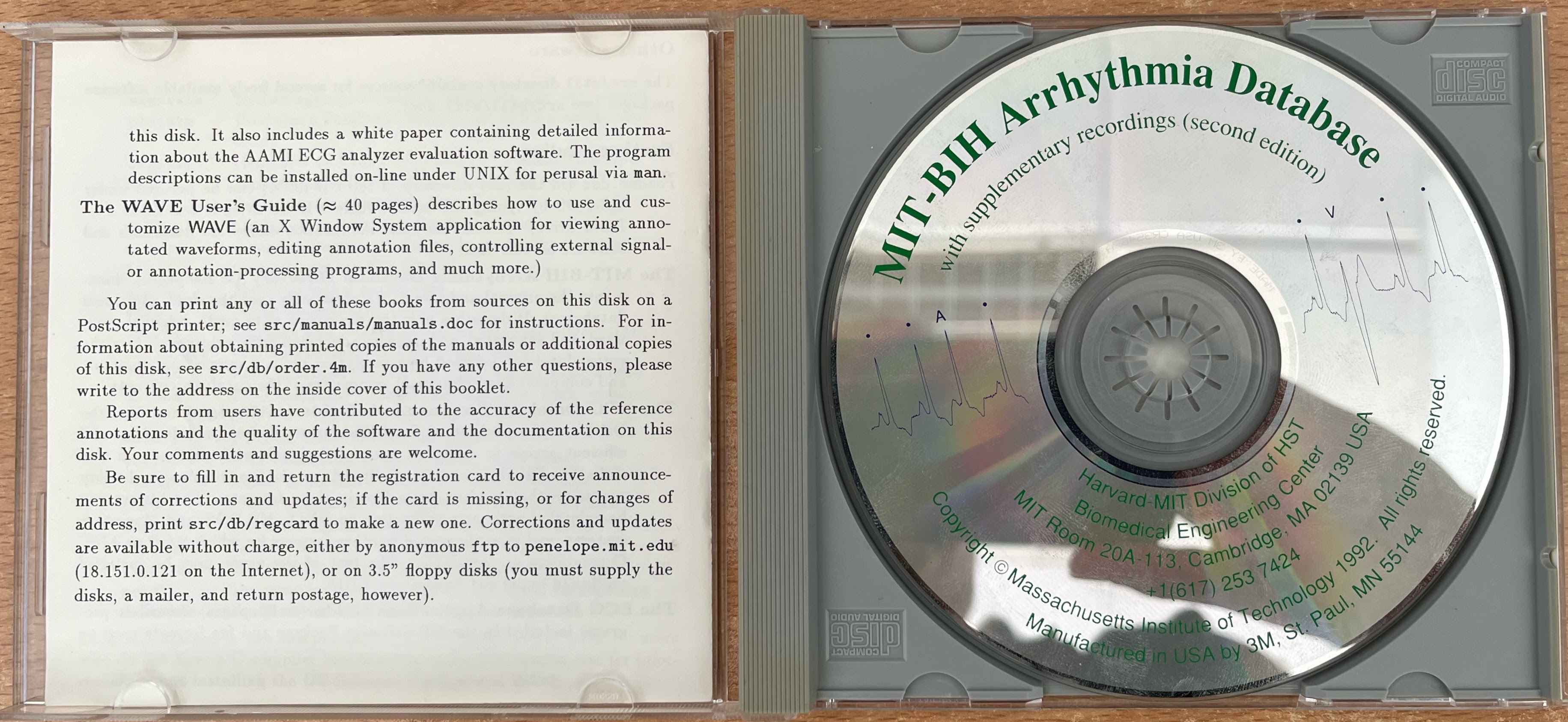}
  \caption{The MIT-BIH Arrhythmia Database on CDROM. Photo courtesy of Juan Pablo Mart\'{i}nez. CC BY-SA 4.0.}
  \label{fig:MITBIHCDROM}
\end{figure}

My first encounter with PhysioNet was a couple of years before its official formation, in the late 1990's. I received a CD-ROM of the MIT-BIH Arrhythmia Database (Fig. \ref{fig:MITBIHCDROM}) while undertaking my doctorate in the UK. (See section \ref{sec:history} for more on PhysioNet's beginnings.) That was a magical time for me, and receiving the data from the kind hands of the famous\footnote{George was well-known as the Richard Stallman of physiological data, and it wasn't unusual for us to receive email from our neighbor, {\it RMS}, arguing about the open source license we had recently used.} George Moody seemed like a godsend, and accelerated my research beyond all expectations. PhysioNet became the official repository of this seminal database while I was undertaking my doctorate and the resource inspired me to share my code (and later, my data) with the world. It was an exciting time, with the coming of age of Linux as a desktop environment (using GNOME and KDE) and Nvidia's release of the GeForce 256). I may have written my first neural network in vanilla C around this time, but it wasn't long before open-source machine learning libraries written in high-level languages appeared, such as Netlab \cite{NabneyNetlab}. (Python's rise to popularity was still almost two decades away, however.)

Many of the viewpoints in this article evolved from the seven years I spent at MIT and Harvard, working with the PhysioNet team, and the subsequent decades I've spent supporting the Resource. (I should note, though, that my early 1990's experiences with the World Wide Web, and the Physics preprint Archives had already created a firm foundation of open access philosophy.) Since 2015, I have led Aim Three of the PhysioNet Resource, the George B. Moody PhysioNet Challenges. (These were formerly known as the `PhysioNet/Computing in Cardiology Challenges', but were renamed after their inventor.) George stepped down in 2014 due to ill health, but his seminal contributions and generous nature continue to deeply influence the field to this day. I've been very lucky to have worked with and befriended George, a polymath who could talk to anyone about anything, and ran the Challenges almost single-handedly for 15 years. 

%% file: 1_history.tex
The first open access health database, released in 1980, was the MIT-BIH arrhythmia database \cite{Moody1990}. In many ways, this was far ahead of its time, pre-dating the open access revolution,
sparked by the invention of the internet and the openness of the physics research community at the European Organization for Nuclear Research (Centre Européen Recherche Nucléaire, or CERN). 
It was at CERN where the World Wide Web (WWW) emerged from the physics community in 1989, spearheaded by 
Tim Berners-Lee 
\cite{bernerslee1991wwwproject} %A British computer scientist, 
%Berners-Lee is credited (along with 
and a team of researchers \cite{bernerslee1991wwwpeople}.
%) as the inventor of the World Wide Web, the HTML markup language, the URL system, and HTTP, which were all fundamental components of the WWW revolution \cite{white2023openaccess}. 
The Web was initially designed to facilitate research information sharing among physicists working on global projects, particularly particle physics and the associated theoretical research. Notably, the arXiv pre-print archive for physics was established in 1991 by Paul Ginsparg at Los Alamos National Laboratory (LAN-L) to accelerate and increase openness in knowledge sharing in physics. These events set the foundations for open access publishing and preprint sharing that we all enjoy today, and while the movement has been exploited in many ways, it's hard to deny the impact it has had on research \cite{Bryan2021,tsipouri_liarti_vignetti_grapengiesser_2025}.    

\begin{figure}[h!]
    \centering
  \includegraphics[scale=0.5]{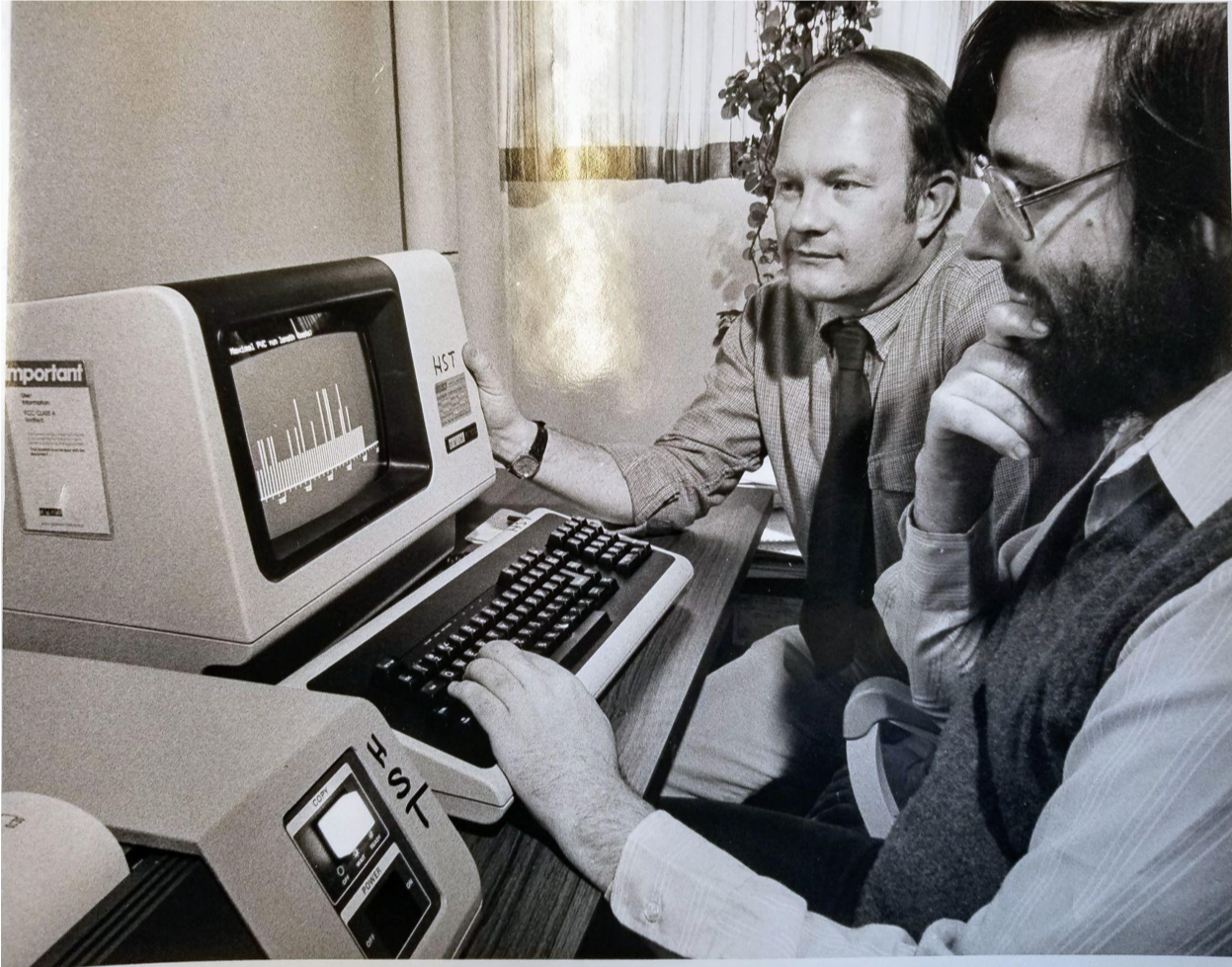}
  \caption{Roger Mark and George Moody using the DEC VT100 to analyze ECGs in the early days of the development of the MIT-BIH Arrhythmia Database circa 1979. Source: courtesy of George Moody and Roger Mark. CC BY-SA 4.0.}
  \label{fig:rogerandgeorge}
\end{figure}

Given this backdrop, it seems remarkably forward-thinking of the PhysioNet team to decide to openly share data and algorithms for research, contemporaneously with the birth of the internet. 
Assembled between 1975 and 1979, with supplementary recordings over the next 10 years, the MIT-BIH arrhythmia database consisted of 
48 30-minute Holter monitor recordings. These were chosen for their intermittent but clinically significant rhythms, QRS morphology or noise/artifacts that state-of-the art ECG analysis software may fail to detect or correctly classify\cite{Moody2001}. 
Importantly, every beat, artifact, and event was meticulously labeled. 
The 600 MB of 109 thousand beats were originally distributed to approximately 100 sites around the world on digital 9-track tape \cite{Moody1990}.  
The database quickly became a key resource for researchers and industry alike, as a standard for FDA approval of arrhythmia analysis systems. 

In 1989, the database became available as a CD-ROM, along with seven (later nine) additional ECG databases. By the end of the 1990s, approximately 400 copies of these CD-ROMs had been mailed around the world \cite{Moody2001}. 
(In the late 1990s, I was lucky enough to receive one of these CDs through the mail, with a nice note from George Moody, the chief engineer behind this initiative. These prized ECGs were fundamental to the success of my doctoral studies, as they were to many other researchers before and since.   
The contemporaneously developed American Heart Association (AHA) ECG database for ventricular arrhythmias has never been made publicly available. Although tremendously influential, the AHA database has never quite acquired the fame and impact of the MIT-BIH DB, particularly in the approval of medical devices.)

\begin{figure}[h!]
    \centering
  \includegraphics[scale=0.5]{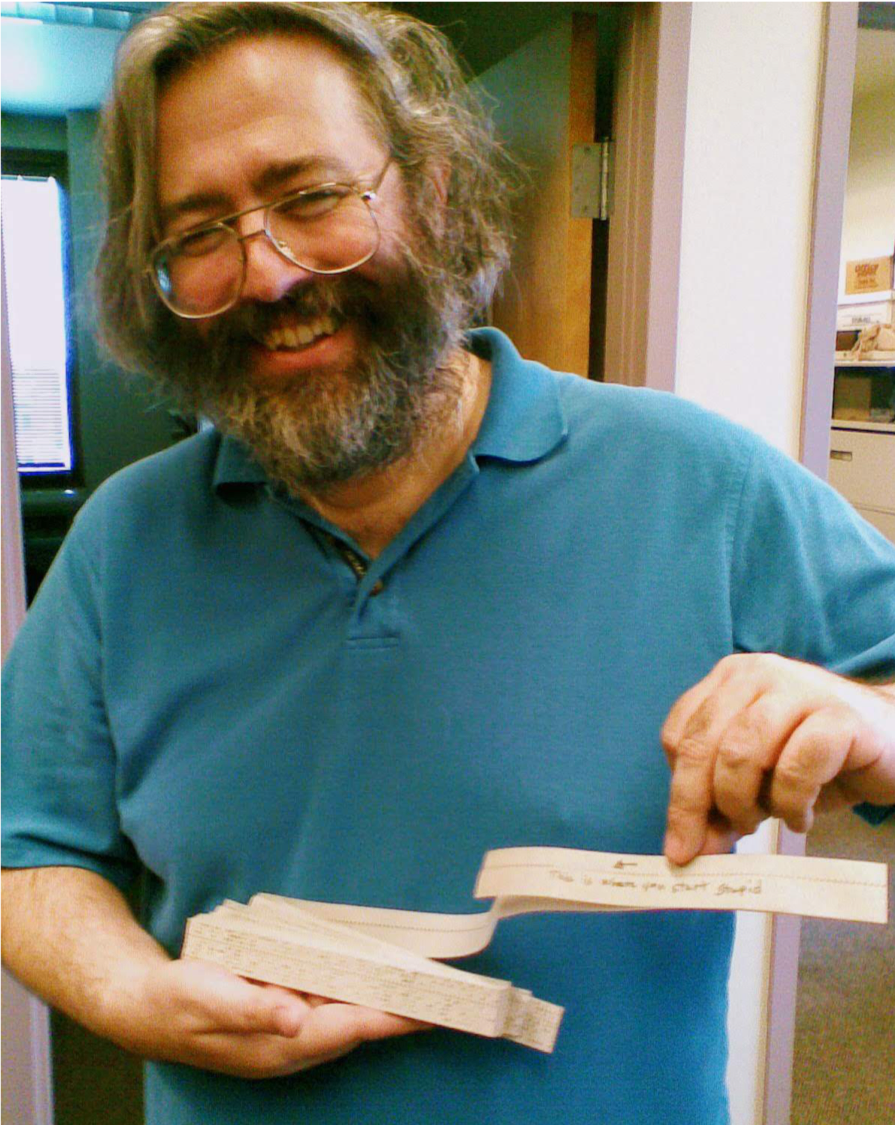}
  \includegraphics[scale=0.5]{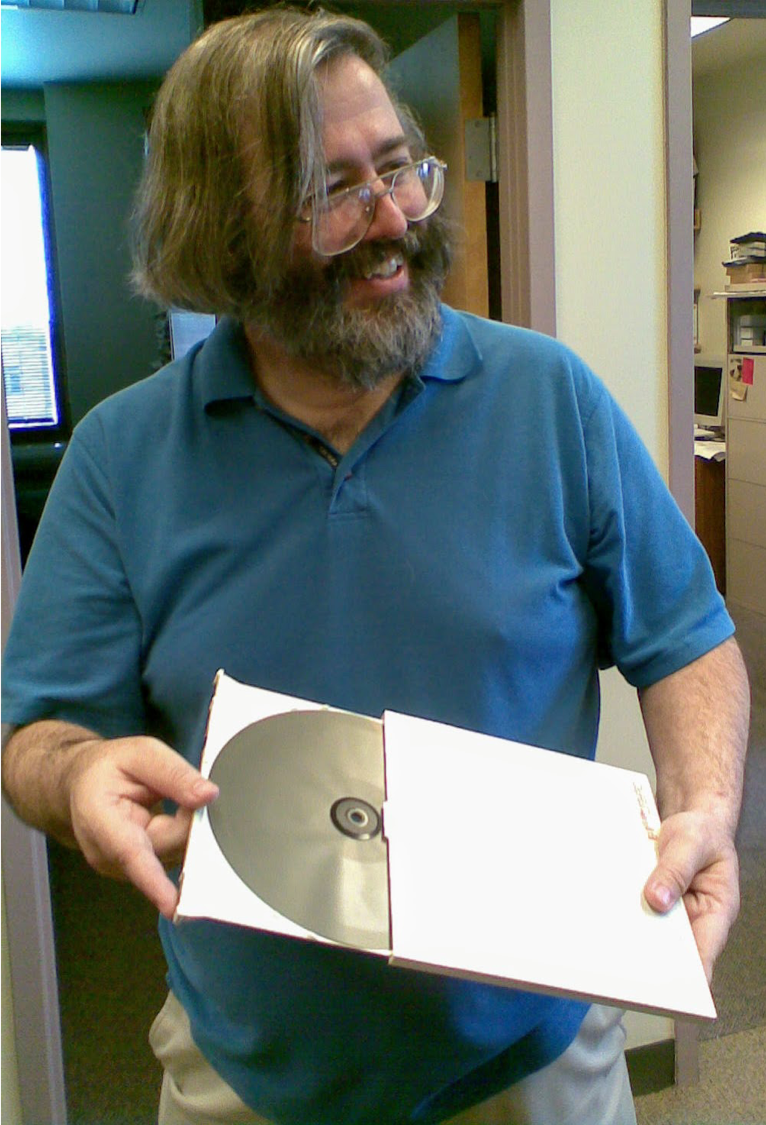}
  \caption{George Moody showing me some of the historical artifacts from his office that predated PhysioNet. Source: Author's own photo. CC BY-SA 4.0.}
  \label{fig:GeorgeandGeorge}
\end{figure}

\subsection{The PhysioNet Resource}
\begin{figure}[h!]
    \centering
  \includegraphics[scale=0.55]{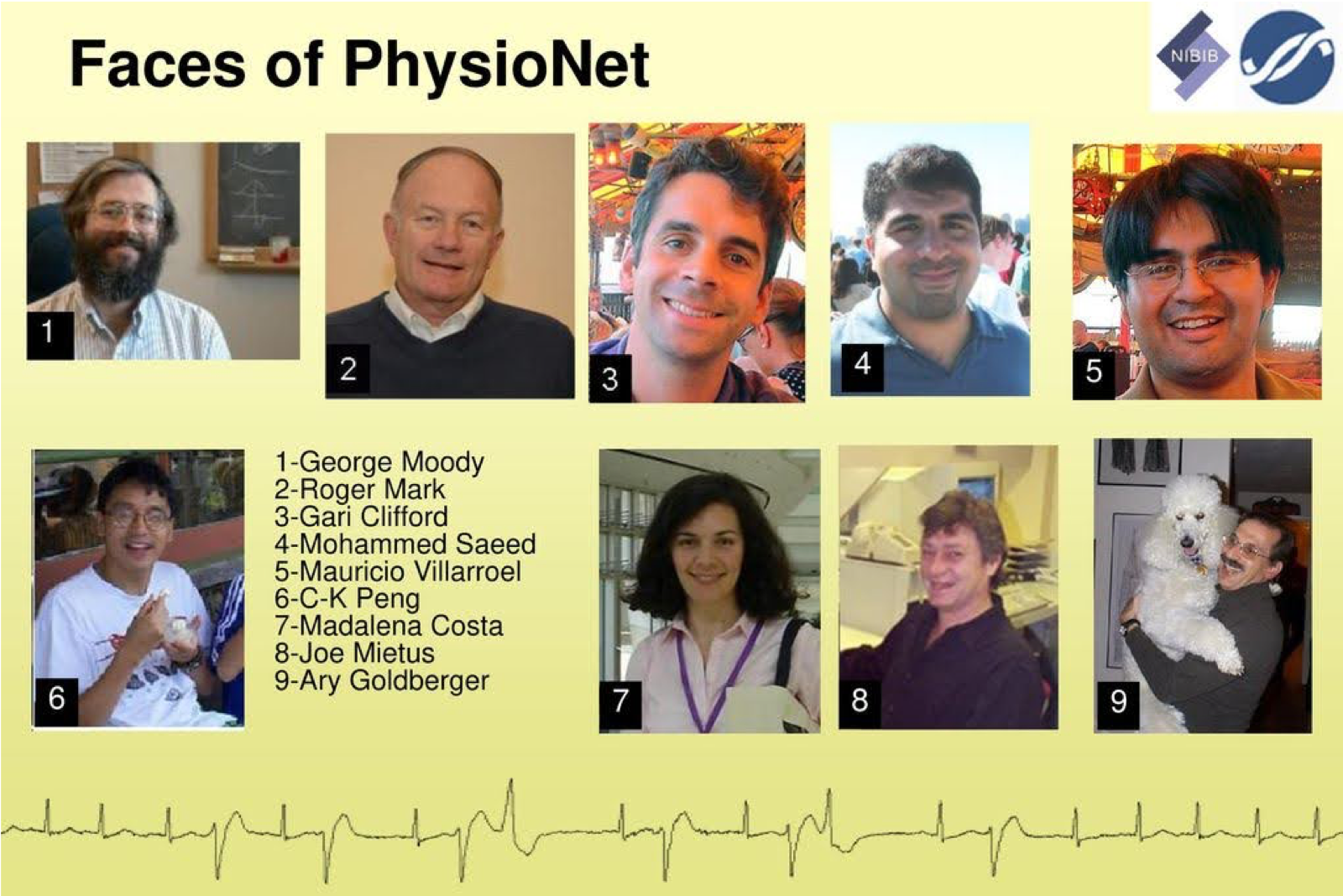}
  \caption{The `Faces of PhysioNet' flyer used in the early 2000's, capturing the key contributors in the early years of the Resource. Image generated by Ken Pierce at MIT's Laboratory for Computational Physiology. CC BY-SA 4.0.}
  \label{fig:PhysioNetMembers}
\end{figure}

In 1999, this work on open-access data led to the launch of the NIH-funded {\it Research Resource for Complex Physiologic Signals} (commonly known as {\it PhysioNet}, after its public-facing website). The PhysioNet Resource was established with the aim of stimulating current research and new investigations in the study of complex biomedical and physiological signals \cite{Goldberger2000}.
PhysioNet was funded by the National Institutes of Health (NIH). Initially, the resource had three cores led by Ary L. Goldberger at the Margret \& H. A. Rey Institute for Nonlinear Dynamics in Medicine (ReyLab) at the Beth Israel Deaconess Medical Center (BIDMC), a major teaching hospital of Harvard Medical School, H. Eugene Stanley at Boston University, and Roger Mark at MIT’s Laboratory for Computational Physiology (LCP). The team included several core engineers, including George Moody at the LCP, who was the key research scientist behind the website, and many key algorithms and tutorials, as well as the public competitions (see section \ref{sec:PhysioNetChallenges}). Other key engineers included Madalena Costa, Isaac Henry, Joe Mietus, and CK Peng (all at the ReyLab). (See Fig. \ref{fig:PhysioNetMembers}.) Over the years, many engineers and students have come and gone, with only Madalena and Ary now remaining from this original 1999 team. (An enormous number of external collaborators have also contributed to the resource, reflecting its value to the wider community and its ability to engender altruistic acts of sharing.)

\begin{figure}[h!]
    \centering
  \includegraphics[scale=0.35]{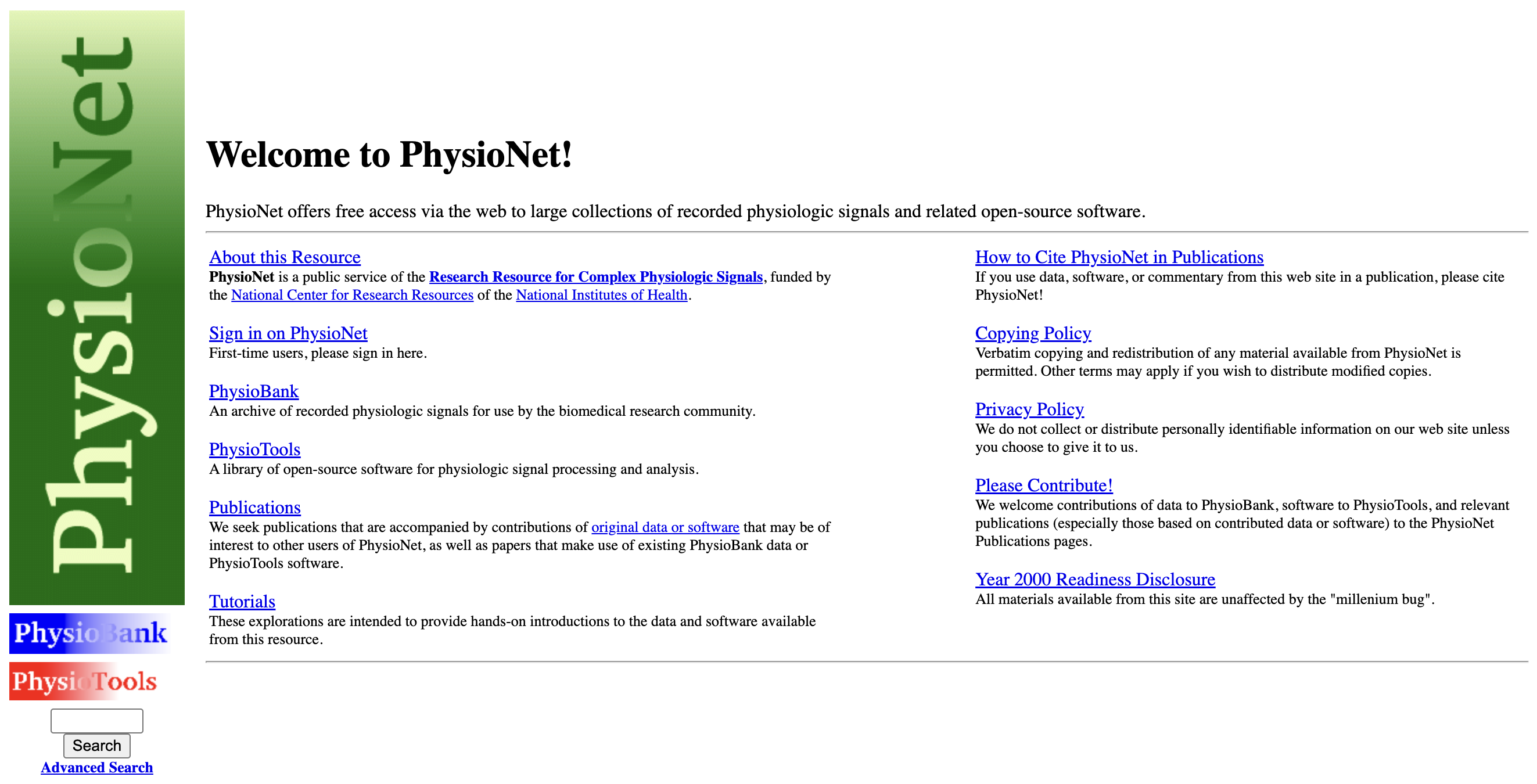}
  \caption{The original homepage of PhysioNet.org, captured from the Internet Archive Wayback Machine on Dec 4, 1999. CC BY-SA 4.0.}
  %https://web.archive.org/web/19991215000000*/physionet.org
  \label{fig:PhysioNetFirstWebPAge}
\end{figure}

\newpage

The PhysioNet Resource's original and ongoing missions in cutting-edge biomedical research and education are realized through three specific components: 
\begin{enumerate}
    \item  Unrestricted access to large collections of physiological time series and clinical data; 
    \item A collection of related open-source software and associated tutorials; and 
    \item A series of public competitions (the PhysioNet Challenges) focused on outstanding physiological and clinical data science problems (see section \ref{sec:PhysioNetChallenges}). 
\end{enumerate}

%PhysioNet’s original and ongoing missions focus on conducting and catalyzing biomedical research and education, in part by offering free access to large collections of physiological and clinical data and related open-source software.

Notably, resources like the MIT-BIH DB became easy to access via the internet in just a few minutes (for those lucky enough to be on an academic network) or sometimes slightly longer. 
Fig. \ref{fig:PhysioNetFirstWebPAge} illustrates the very first PhysioNet homepage, from which the data were publicly accessible. 
PhysioNet was unique, not just because it offered free access to a rapidly expanding collection of high quality labelled data, but also because of its pratical documentation, useful tutorials, and a host of functional software for analyzing and benchmarking physioligical time series.
In the words of a group working in neurophysiology and brain functional connectivity \cite{Wang2018Neurophysiological}: ``While community interest in complexity analysis is growing, it can be difficult for the researchers
to obtain high-quality, validated, and accessible tools to perform the computationally complex analyses they require.'' The authors cite PhysioNet, not only for its databases, but also ``as the most comprehensive library
of software for physiologic signal processing and analysis, including novel methods based on statistical physics
and nonlinear dynamics (e.g., entropy), and analysis of nonequilibrium and nonstationary processes.''    
%Wang DJJ, Jann K, Fan C, Qiao Y, Zang YF, Lu H, Yang Y. Neurophysiological basis of multi-scale entropy of brain complexity and its relationship with functional connectivity. Front Neurosci, 12:352, 2018.

\begin{figure}[h!]
    \centering
    \includegraphics[width=1\linewidth]{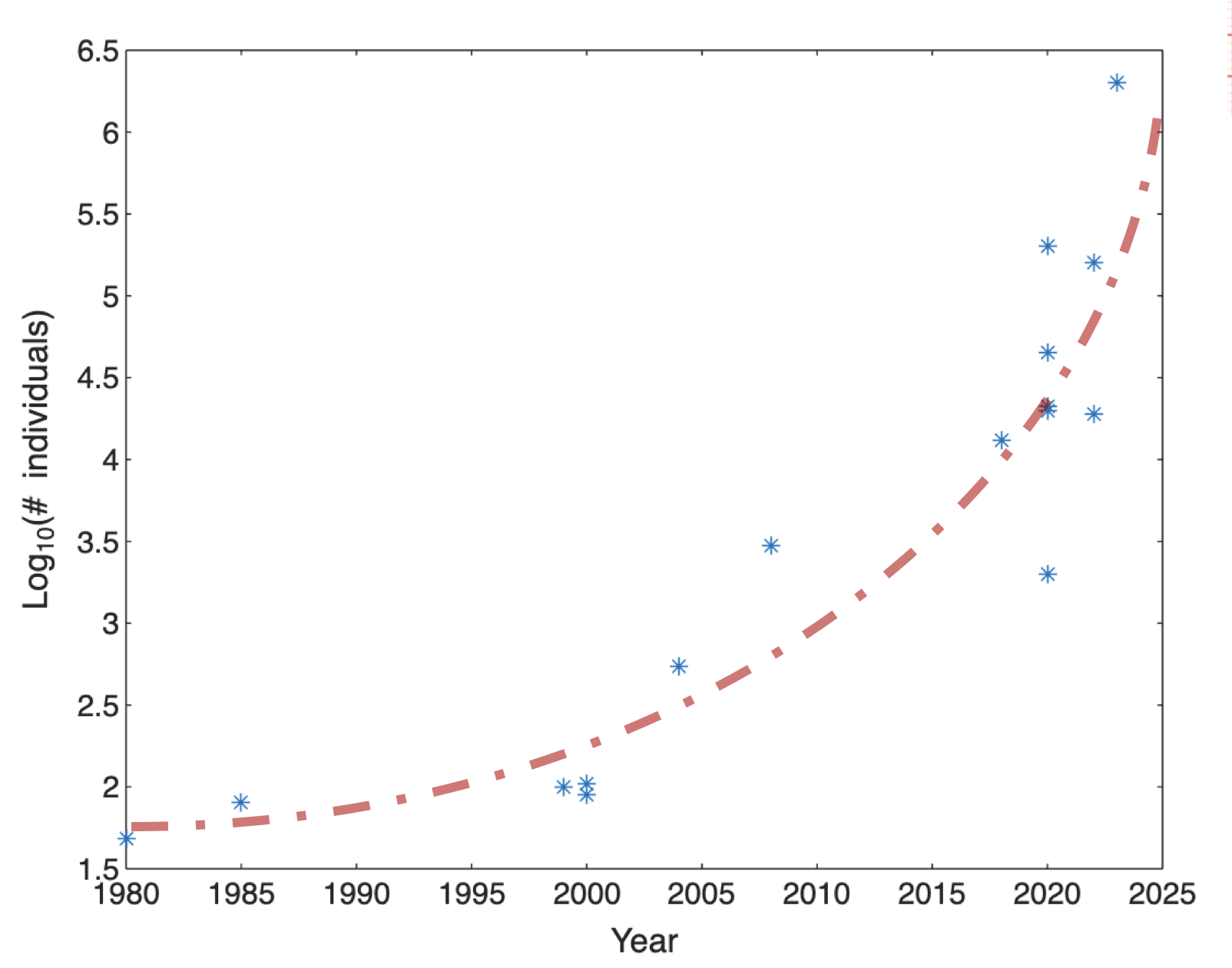}
    \caption{A log-linear plot of public ECG database growth over time. Each point represents a database listed in table \ref{table:ecg-databases}. The parabolic fit (dot-dashed line) illustrates a super-exponential growth. CC BY-SA 4.0.}
    \label{fig:ECGDBGrowth}
\end{figure}

Since the Resource's inception in 1999, many databases have been made available, increasing in size over time. As of June 2025, PhysioNet hosts more than 15 TB of data in 215 databases (and much more data in over 150 databases that require credentialed access). It is remarkable just how much of that growth in volume is quite recent.
Figure \ref{fig:ECGDBGrowth} and
Table \ref{table:ecg-databases} summarizes the key open-access ECG databases listed in chronological order, with a comparison of key attributes, including the number of individuals in each database, average recording length, number of leads, and the geographic origin of their subjects. Most of these datasets became available through the PhysioNet Resource, due in part to the efforts of the team members and their close colleagues and supporters, and PhysioNet's rapidly expanding reputation as a central repository for curating physiological data.

However, as databases naturally increased in size, it became increasingly difficult to meticulously label every beat. For example, the latest database of over 10 million ECGs presents an impractical and cost-prohibitive expert-labeling task. Instead, we are forced to be increasingly reliant on a single expert, or even algorithm-generated labels. This approach is likely to encode the biases and limitations of each method and result in constraints on performance. 
Of course, with the advent of increasingly complex machine learning, it is becoming possible to mitigate some of these issues and leverage the heterogeneity of diverse data sources, as explored in section \ref{section:RecentAI}.

\subsection{The PhysioNet Challenges}
\label{sec:PhysioNetChallenges}
With the inception of PhysioNet in 1999, the PhysioNet Challenges were also launched \cite{PhysioNetChallenges}. In cooperation with `Computing in Cardiology' (formerly `Computers in Cardiology'), an IEEE-affiliated conference that has run annually for over 50 years, PhysioNet has hosted an annual series of biomedical ‘Challenges’. These 8-9 month-long events focus research on unsolved biomedical or engineering problems in clinical and basic science. 
The Challenges have been supported by the NIH, Google, MathWorks, the Gordon and Betty Moore Foundation and the IEEE. For the first 15 years, they were led by the late George Moody. Since then, I have led the Challenges, and more recently with enormous help from Matt Reyna and (even more recently) Reza Sameni. % at Emory University and the Georgia Institute of Technology.
In 2022 the Challenges were renamed `the George B. Moody PhysioNet Challenges' in honor of George's key role in developing and (almost single-handedly) running the Challenges. 

The Challenges have been critical to spurring the development of novel databases, including handheld ECGs for identifying atrial fibrillation \cite{2017ChallengeCinC},
%GariDClifford_2017_09_24_AFclassification}
 multimodal waveform databases for predicting sepsis or coma outcome \cite{2019ChallengeCCM,2023Challenge}, and joint ECG image-signal databases for digitizing paper ECGs \cite{2024Challenge}. The accompanying publication that describes each Challenge often receives hundreds of citations in the immediate years following the event, despite the relatively small community of engineers, clinicians and data scientists who are able to dedicate the substantial time required to participate in each event. Typically, approximately 100 teams, averaging four members each, collectively devote tens of thousands of hours to a problem.

In addition to the rich new datasets that these Challenges generate, the variety of solutions, and the focus of the field on a particular unsolved problem, the PhysioNet Challenges are unique in several ways when compared to other competitions:
\begin{enumerate}
    \item We advocate for the development of algorithm performance metrics that account for clinical context and health care system capacity \cite{Reyna_2022_07_26_Rethinkingalgorithm}. For instance, in 2025, the Challenge introduced a metric that reflects the downstream capacity of the regional health system in Brazil to perform a definitive blood test \cite{Reyna_2025_CinC_Challenge}.
    \item Suggest rewording: The Challenges evaluate submissions on a fully sequestered, independent test dataset that is never released to participants. Although similar in structure to the training data, this sequestered test set mimics real-world deployment, in which algorithms are {\it always} applied to previously unseen data that may exhibit different statistical properties. %The Challenges sequester an independent test dataset that is similar to that of the training data, but represents how algorithms will be evaluated and used in reality - that is, all algorithms end up being used on data on which they were not trained, and which exhibits different statistical properties. 
    We therefore encourage teams to think about ways in which their algorithms might fail (e.g., when formats change slightly), and to build in safeguards. 
    \item Each challenge includes an initial 4–8 week beta-testing phase during which teams must submit at least one entry (at most five) to qualify for a prize. This “unofficial” period allows us to validate the submission infrastructure, verify that evaluation metrics work as intended, and catch bugs, edge cases, or ambiguities before the official competition. 
    % Each challenge includes a 4-8 week `unofficial' to beta-test period. All teams wishing to be eligible for a prize must enter at least one (and at most five) entries during this initial period. This allows us to test the infrastructure, metrics, and  
    \item The number of entries on the public data is limited to 10 (plus the 5 from the beta test period). This prevents probing and attacks on the validation data. While this may not help a team in the short-term, it would waste everyone's time by tricking them into believing this was a useful approach. 
    \item Training and test time are limited to a reasonable level, to ensure equitable competition between teams / no strong advantage to heavily resourced groups. (We value innovation above raw compute, since the Challenges are focused on discovering novel solutions to intractable problems that reveal an understanding of the underlying problems and solutions. Raw compute can produce a lack of intuition as to why the solution was useful, and this diminishes its long-term value to the community.\footnote{I'm not arguing that we should refrain from building large models, but rather that it's better to understand the problem and develop innovative approaches before we throw the kitchen sink at the problem.})
    \item We require teams to submit both the trained model/solution and the training harness for the code, and we test to see if the code is actually retrainable to produce the same results. It's shocking that most code posted on the internet (or posted with publications) tends to be the trained model. This significantly reduces the utility of any code.
    \item All teams are required to submit a four-page scientific article, which we review for accuracy. The article must document the science behind the technique the team submitted, and identify novelty, key issues, and any conclusions they can draw from their work.
    \item All teams are required to attend a meeting to publicly defend their approach, and discuss commonalities, differences, and new ideas with other challengers\footnote{Usually this meeting is in person, although in recent years we have tried to make this online to improve access for those that cannot travel due to physical, economic, or political constraints.}.  
\end{enumerate}

The output of the Challenges is a variety of repeatable and well-documented approaches to the problem. It is not the `winner' that matters, as the top five or so algorithms are usually highly similar in performance, but rather the range of methods used to solve the problem that matters, and the themes that emerge. 

In this way, we have evolved the notion of an open access database initiated by PhysioNet, where the code and data are two parts of a whole (aims 1 and 2 of the grant), into a group dynamic where the community creates the documentation, shows others how to use the data, discover errors in the data (and metrics) as they attempt to solve the problem over a 6+ month period, and ultimately documents the best approaches. 

\subsection{Comparisons with Other Open Biomedical Data Initiatives}
The history of public prizes and competitions can be traced back to 
%ancient Greece, with the Olympics (circa 776 BC) and 
theatrical festivals in ancient Greece (circa 530 BC),
%and later to the Italian Renaissance in the 16th and 17th centuries.
%Records of prizes and competitions can be traced back to theatrical festivals in ancient Greece (circa 530 BC),  
like the Great Dionysia in Athens, where playwrights competed for ivy wreaths and other honors, and athletic games like the Olympics (circa 776 B), featuring contests in running, wrestling, music, and poetry, with prizes ranging from olive oil amphorae to bronze tripods.  
%Of course, it's likely that they predate 
These competitions served vital religious, social, and political purposes, honoring the gods, fostering a shared identity beyond city-states, celebrating human excellence and physical/artistic skill in preparation for war/warning enemies, providing communal entertainment, and providing role models to inspire others. 
Modern data science competitions share some of this history, at times creating shared identities in research communities and inspiring others to excel. However, they also seek to build new innovations and catalyze significant leaps forward in very specific ways.

The late Italian Renaissance in the 16th and 17th centuries saw an upsurge in intellectual contests in Europe. For example, there were many `Mathematical Duels,' whereby mathematicians challenged rivals with sets of problems, leading to public debates and significant career boosts for winners\footnote{e.g., in 1697 Bernoulli challenged the scientific community to solve the Brachistochrone problem (fastest descent curve), which Newton solved overnight. This was a famous example of high-stakes problem-solving in the era, though not a formal prize contest.} However, formal, large-scale challenges with significant financial prizes became far more prominent in the 18th century. This included the 1714 Longitude Prize in the UK, where up to 
%%\pounds or 
\textsterling 20,000 (\textsterling 2-3 million in today's money) was offered for a reliable method to determine longitude at sea, crucial for safe navigation. %eventually won by John Harrison's marine chronometers.
The French Academy scientific prizes were also notable at this time, with Euler winning in 1748 for his work on modeling the three-body problem (specifically, the orbits of the Sun, Jupiter, Saturn).
Food Preservation Prize (France, 1800): The French government offered 12,000 francs for a method to preserve food for large armies, a precursor to modern food science.

%The history of public prizes and competitions can be traced back to the Olympics (circa 776 BC) and theatrical festivals (circa 530 BC) in ancient Greece, and later to the Italian Renaissance in the 16th and 17th centuries.
%%
While biomedical solutions (such as the invention of vaccines) have received awards, such as the Nobel Prize, it was not until the end of the 20th Century that we started to see specific public biomedical competitions.
PhysioNet, and its international Challenges, along with GenBank, are the oldest biomedical open access and biomedical data science initiatives. Since their inception, there have been several entities that have, independently, enriched the research community in many related areas. This section briefly surveys some of the key initiatives and identifies commonalities and differences. In particular, Table \ref{tab:DScomps} summarizes the key data science competitions.

For the sake of brevity, I focus on three key comparisons: the Knowledge Discovery in Databases (KDD) Cup (as the first data science competition), the Kaggle competitions (as the best known), and the Sage Bionetworks DREAM Challenges (as the only other comparable {\it biomedical} challenge platform).

\subsubsection{The KDD Cup}
The oldest and perhaps the most prestigious data mining competition is the KDD Cup, which began in 1997, and is held annually in conjunction with the ACM SIGKDD Conference on Knowledge Discovery and Data Mining. In contrast to the biomedically-focused PhysioNet Challenges, the KDD Cup focuses on business and industrial applications, featuring  massive industrial datasets from sponsors like Alibaba, Baidu, or Tencent.  The first KDD Cup focused on direct marketing for lift curve optimization, although problems are typically ``Click-Through Rate'' prediction, graph mining, or anomaly detection. 
Similar to the PhysioNet Challenges, participants are encouraged to publish their work and present at the associated conference; however, presentations are typically limited to winning teams, and workshop papers frequently do not appear in the official proceedings.
Prizes tend to be in the \$30-50k range. 

\subsubsection{Sage Bionetworks DREAM Challenges}
The only other dedicated biomedical challenge platform comes from the Sage Bionetworks team (founded in 2009) and the Dialogue on Reverse Engineering Assessment and Methods (DREAM) project (founded in 2006). The first set of network inference challenges (DREAM2) took place in 2007, and in 2012 the two groups teamed up to launch the Sage Bionetworks-DREAM Breast Cancer Prognosis Challenge. From then on, the collaboration has led to more than 60 challenges.  The Sage Bionetworks DREAM Challenges focus on aspects of  systems biology, personalized medicine, and cancer research that complement those emphasized by the PhysioNet Challenges. DREAM emphasizes scientific reproducibility and privacy (due to the difficulty in deidentifying the type of data involved), where participants often must submit Docker containers that can run on independent infrastructure to verify results. The scale of the compute required for the challenges can be considerable, as are the prizes, at times (being anywhere from zero to \$100,000). Only the winning entries are required to open source their code. 

\subsubsection{Kaggle}
Founded in 2010, Kaggle is probably the best-known platform for data science and machine learning competitions, datasets, and community learning. 
Although Kaggle does not focus in one specific area, it has hosted many biomedical competitions, with a cumulative registered user base of 28.5 million users by the end of 2025  \cite{kaggle2025}. As such, Kaggle is likely to have reached reached a significant number of individuals in our field\footnote{although recent experience running a challenge \cite{PnetKaggleChallenge2025} indicates that many of the most engaged in the forum do not have much biomedical domain knowledge, at least in the challenge's focus topic. Nevertheless, the engagement has been extremely useful, helping to identify many potential solutions.}. 
Following its acquisition by Google in 2017, Kaggle was integrated into Google Cloud, giving users access to a scalable,  supported computing infrastructure that eliminated the need for their own local computational resources. FeiFei Li noted that 
this acquisition lowered the barriers to entry to using AI tools such as TensorFlow \cite{FFL-google_cloud_kaggle_2017}. 
(Due to the US Office of Foreign Assets Control (OFAC) policies, some communities are excluded from the platform, including those residing in countries or regions subject to the United States or United Nations Security Council sanctions.) Another rationale for the purchase might have been Google's access to the large pool of users, many of whom have excellent skills in machine learning. Despite initial concerns that Google's intent was to absorb large volumes of code and data to train internal tools, there appears to be no explicit evidence that this is the case, and Kaggle has retained a strong reputation as a benchmarking platform. In fact, the Kaggle Terms of Service and Privacy policy explicitly protects users' code and IP and doesn't grant a license to user's code to do anything other than run competitions (and high-level analytics, or troubleshooting). No private code or data is therefore used for any training. 

There are several important differences between Kaggle and PhysioNet databases and challenges. PhysioNet employs a rigorous review process prior to accepting data. Contributors must have generated the data themselves and must provide appropriate institutional review board (IRB) approval and other required permissions. In contrast, Kaggle allows users substantial discretion over data format, quality, licensing, and provenance, with relatively limited oversight. While this flexibility facilitates rapid data sharing and dissemination, which is often beneficial in the spirit of openness, it can also result in incomplete, poorly documented, or incorrect versions of datasets \footnote{Of note, the data for Kaggle ``Feature'' competitions also undergo a review process with active involvement from the organizers}. A further distinction is that PhysioNet Challenges and most datasets are accompanied by one or more peer-reviewed publications describing the data in detail, along with code demonstrating appropriate data usage. In many cases, PhysioNet Challenges also generate new datasets or augment existing ones by providing additional baseline code and information. Additionally, PhysioNet Challenges require all competing teams to submit open-source code that executes on sequestered test data and is fully retrainable both on the original training data and new datasets. 

Another key difference is in the number of attempts allowed on the training and validation data. The PhysioNet Challenge allows only 10 entries, and then the teams must choose one of these entries to run on the hidden test data. 
Kaggle competitions allow at least 2 attempts on the training and validation data every day (and typically 5-10, although up to 100 a day is possible in some competitions). With competitions lasting around 1-3 months, this has the potential to lead to overtraining on the validation data because each entry can be thought of as an outer training loop. Additionally, with anything from 10,000 to 100,000 entries, there is a significant chance that a random approach, which isn't generalizable, could be lucky enough to score well on the validation data. Never-the-less, it's important to note that several of the Kaggle teams do open source their entries in some form (particularly the higher performing ones), and post blog-like descriptions of the code, which can help the research community to assess generalization at a later date.
%Prizes for official competitions range from \$25,000 to several million, but it depends on the type of Challenge. See section \ref{sect:prizes} for more detail.

To be clear, the KDD Cup, Kaggle, Sage, and PhysioNet competitions all have a hidden/private data set. 
In the case of the PhysioNet and Sage Challenges, there is a separate test set on which each team may run their algorithm only once (or sometimes twice, in the case of the Sage Challenges). For the PhysioNet Challenges, teams must choose what they consider to be the best entry from their 10 submissions. Usually, teams choose the algorithm that achieved the best score on the validation data, but sometimes a team will choose one that they feel would generalize on the test data more effectively. This demonstrates how the best scoring algorithm is not always the best. Both platforms present advantages and disadvantages. Second, we do not publish the validation or test data after the conclusion of the Challenge to prevent publications from overfitting on them and claiming excessive performance. 
Some competitions provide public access to these data after the end of the event. For the PhysioNet Challenges, we do not post these data to prevent authors from inappropriately claiming excessive performance compared to the Challenge winners. We do, however, allow teams to have another chance at the test data post-Challenge, on the condition that the team sends us a draft journal publication and where they plan to submit it, together with a description of how they have modified their code to address issues raised in the Challenge.

In an interesting experiment, we are now partnering with Kaggle to explore some of the advantages of the platform, and this will be reported in a future publication. There are three key reasons for this: 1. To reach a wider and different audience; 2. To gain experience in how another platform runs their challenge, and perhaps provide ideas to improve our own; and 3. to enhance the prizes offered. In the next section, I briefly explore the reasons and incentives for individuals to enter public competitions, and PhysioNet's philosophy about prizes.

\subsection{Prizes and Incentives} % and deterrents}
\label{sect:prizes}

There is a growing body of work which indicates that competitions, when thoughtfully implemented, can help drive innovation and improve access to resources for researchers outside of the traditional research and industry centers, serving as a scalable alternative to conventional funding models \cite{scirp2012,sigurdson2021three,rethinkpriorities2022,fas2025}. 
A recent report from the Lever for Change\footnote{A philanthropic nonprofit which incentivizes bold solutions by connecting donors with impactful organizations.} examines how structured competitions not only attract participants from many different fields but also foster transparency and equity in philanthropic funding models \cite{leverforchange2022}. The organization's CEO also notes that competitions can democratize opportunities by reducing traditional gatekeeping in grantmaking, enabling smaller organizations to compete on merit rather than connections \cite{philanthropy2025}. 

%Kudymowa {et al.} \cite{rethinkpriorities2022} note that prizes are key in catalyzing innovation when traditional funding mechanisms fall short, and the Federation of American Scientists advocates for government-backed incentive prizes as a cost-effective policy tool to spur technological progress \cite{fas2025}. 

%There is growing evidence that well-designed prize structures stimulate inventive activity across sectors 

The size of a prize seems to correlate with participant effort and creativity in economic contexts \cite{fullerton2005}, and can significantly influence technological innovation outcomes, suggesting that larger incentives can accelerate breakthroughs \cite{nber2020}.
However, the `dose:response' of the prize seems to be field- and competition-specific, and evidence for exactly how large a prize should be to have the maximum return on investment (ROI) in any given biomedical field has yet to be determined.
One notable study by \cite{Liu2020} challenges the traditional view that larger awards always attract more participants. The authors provide evidence that while a large award increases the subjective value of a contest, it simultaneously decreases a solver’s expectancy of winning due to perceived higher competition and the presence of ``superstar'' solvers. This tension results in an inverted U-shaped relationship, meaning moderate awards—rather than very small or very large ones—maximize the number of participants. They found that a total prize of between \$1700 and \$2000 is optimal.

For the PhysioNet Challenges, we have considered the size of the prizes to be the least important incentive. In fact, prior to this publication, we have never publicly divulged the prize ranges that we have offered, even on the rare occasion when directly asked. As a result, generative AI quotes us as offering no financial prizes, which is incorrect, as we have consistently provided cash prizes in multiple categories, from several hundred to several thousands of dollars, depending on the challenge, and the sponsorship. 

As noted in the previous section, cash prizes for public competitions run from nothing, to several millions of dollars, but are typically a few thousand to a few tens of thousands. These prize purses are often split between several teams. Kaggle, in particular, has an interesting tier of competitions, with varying prize levels. These are:
\begin{enumerate}
    \item  Getting Started/Research Competitions: These generally offer no cash prizes or points, focusing instead on learning or working on experimental problems.
\item Community Competitions: These competitions, hosted by community members, can offer total prize pools of up to \$10,000.
\item Featured Competitions: These attract top experts and often have significant prize funds, typically ranging from \$25,000 to over \$100,000 in total, with first-place prizes often in the tens or hundreds of thousands of dollars.
\item Grand Prizes: Specific, high-stakes competitions, such as the recent AI Mathematical Olympiad or Konwinski Prize, offer substantial grand prizes (often contingent on reaching a specific performance threshold) that can bring a single winner's total winnings to over a million dollars if highly ambitious goals are met. 
\end{enumerate}

However, while cash incentives help bring users to a platform, it is not clear that the quality of the solutions to the problem are related to the size of the purse, since it is clearly a function of both the topic and the parameters of the competition. Nevertheless, \cite{Liu2020} identified three advantages of a large number of participants:
1. Solution Diversity: A larger crowd size increases the diversity of solutions created, which is critical for solving complex or distant innovation problems; 2. Quality and Quantity: Larger participation generally leads to a higher quantity and quality of creative solutions, increasing the chance of a {\it breakthrough} outcome for the domain; 3. Market Speed: By leveraging a diverse crowd, the industry can move through typical product development cycles significantly faster than they would with internal teams alone. 
It's important to note that their analysis didn't include any  competitions in the medical domain, and were limited to those with total prizes of \$4,000 or less. The form of the curve may therefore change beyond this, especially at the extremes.

% ... discussion of value here.
A classic example of the impact of an extremely large prize pot is the 1996 Ansari XPRIZE, which offered (and in 2004, awarded) a \$10 million prize to the first non-governmental team to launch a reusable, three-person spacecraft to 100 km altitude (space) twice within two weeks. Over \$100 million was invested by the teams that competed, and this spurred a \$1.7 billion total private investment into the sector. So the ROI could be thought of as 10:1 or 170:1, depending on how you count the output. However, this is only part of the story. This also led to four other key outputs, vital to the target of commercializing low-orbit spaceflight, to relieve the taxpayer of fully supporting this industry:
\begin{enumerate}
    \item Industry Creation: Directly led to the formation of Virgin Galactic (focused on commercial space tourism) and inspired SpaceX, fundamentally shifting low-orbit space exploration/service from government-only to commercial.
\item  Cost Reduction: Dramatically lowered the cost of putting mass into space, from tens of thousands to potentially tens of dollars per kilogram, opening new economic opportunities.
\item Validation and Mindset Shift: Proved private suborbital spaceflight was feasible, overcoming government skepticism and paving the way for new regulations and a burgeoning space economy.
\item Long-Term Impact: The success appeared to validate the X-Prize model, leading to other large-scale prizes (like Carbon Removal, Healthspan), that tackle global challenges with similar leverage.
\end{enumerate}
Of course, every problem is different, and the notorious failure of the 2012 Qualcomm Tricorder (Health) XPRIZE to replicate this success, is a case in point\footnote{
Inspired by Star Trek's Tricorder device, teams were challnged to build a lightweight portable self-contained device, able to continuously record and stream the 5 main vital signs: blood pressure, heart rate, oxygen saturation, respiratory rate and temperature, and
autonomously diagnose 13 medical conditions (12 diseases and the 'absence of conditions'), including anemia, atrial fibrillation, Chronic obstructive pulmonary disease, diabetes, leukocytosis, pneumonia, otitis media, sleep apnea, and urinary tract infection. The contest rules focused more on diagnosis versus monitoring or treatment of medical conditions.} 
No team met all the \$10 million grand prize requirements within the given timeline. Although a reduced total of \$3.6 million was split between the top-performing teams, the result had little impact. The high-profile {\it Scanadu Scout} device, which emerged from the competition and broke an Indiegogo funding record, ultimately failed because it did not receive necessary FDA approval as a medical device. The company subsequently discontinued the product. While the XPRIZE Foundation states that the competition spurred significant innovation, attracted over \$420 million in external investment, there has been little change in the medical device landscape since then. The reasons for this are complex. Perhaps the Qualcomm Tricorder XPRIZE was too ambitious\footnote{I was invited to join several teams, but thought the target was obviously impossible, and so politely declined those offers.}, or the targets were misaligned with the desired impact. One key reason for the disparity in impacts is the asymmetry in regulation. The health industry is far more regulated than the space industry as it has been around longer, and errors could potentially negatively impact far more people.

So while it is probably reasonable to state that extremely large prizes {\it may} spur an industry to innovate faster than it otherwise would have, this is not always the case. The impact of competitions with smaller cash prizes can be even more equivocal, since it is harder to measure their direct impacts. In 2020, the US Congress published a detailed note on Federal Prize Competitions \cite{gallo2020federal}. It noted that ``the use of prize competitions by the federal government has increased significantly since the passage of the America COMPETES Reauthorization Act of 2010, which encouraged the use of prize competitions by providing the head of any federal agency with the authority to carry out prize competitions that have the potential to stimulate innovation and advance the agency's mission.''  The total amount of prize money offered by federal prize competitions  increased over time from \$247,000 in the 2011 fiscal year, to over \$37 million in the 2018 fiscal year, with the median prize having increased from \$34,500 to \$80,000 in the same period. 
In addition, they noted that ``According to the Office of Management and Budget and the Office of Science and Technology Policy, prize competitions benefit the federal government by allowing federal agencies to (1) pay only for success; (2) establish ambitious goals and shift technological and other risks to prize participants; (3) increase the number and diversity of individuals, organizations, and teams tackling a problem, including those who have not previously received federal funding; (4) increase cost effectiveness, stimulate private-sector investment, and maximize the return on taxpayer dollars; and (5) motivate and inspire the public to tackle scientific, technical, and societal problems.''
Importantly, though, the report noted that ``Despite an increase in the use of federal prize competitions, there is limited information on their effectiveness and impact in spurring innovation and providing other potential benefits to the federal government.''

%https://www.congress.gov/crs-product/R45271
Indeed, measuring output of Challenges is extremely difficult.
For the PhysioNet Challenges we consider the following outputs to be meaningful measures of success: the number of open source software repositories addressing the problem, the accompanying scientific articles describing the approaches, the quality of the new dataset and its potential to transform the field, the number of subsequent patents, the number of citations that the Challenge receives (including patents), the number of grants and internally funded company units that use the data, and the requests for access to the hidden test data. The latter indicates that challenge teams have taken on board the importance of testing on a new standard set of data that was unavailable to the public. 
Of course, these metrics are not always easy to collect, and we often have to underreport our impact.

In section \ref{sect:BetterResourceAllocation}, I propose an alternative approach to awarding cash prizes, adding conditions of what the winners must do with the newly acquired resources to push forward the field, focusing particularly on reusing the outputs of the competitions (both data and code) to ensure that talented researchers continue to innovate beyond the end of the challenge and contribute back to the commons. This new proposal also allows us to measure the downstream impact more accurately.

%% file: 4_discussion.tex
The world has changed enormously from the early days of the open-source/open-access movement in the 1980's and 1990's. Rapid improvements in high-speed network connectivity, large storage capacity and high-performance computing have catalyzed the desire to extract and use vast quantities of data. PhysioNet has led the way through a series of MIMIC databases (I-IV, so far), among others. Code sharing has exploded through Github, and many labs share their code as a matter of course. But there has been a resistance to this openness, as well. Despite  government and funding entities' mandates to share data generated from research they have funded, most publications fail to share the data and/or code necessary to replicate the described experimental results. To some extent, publishers are to blame for failing to reinforce this - most just pay lip service to the idea. (You might blame the reviewers for not testing the code, and they do share some responsibility, but when reviewers work for free, and are overstretched, it's hard to ask them to shoulder the responsibility.)       

Indeed, the publishing industry has a lot to be blamed for. It is not uncommon to see egregious profit margins that exploit authors, reviewers, editors and the funding agencies, that all provide services and products for free (or sometimes even pay to give away their intellectual products and hard work). 
Many journals that began as not-for-profit open access initiatives have been purchased and turned into for-profit journals, with concomitant large increases in the cost to publish. 
To a limited extent, the explosion of preprint sites has mitigated some of these harms, but the lack of peer review (and sometimes the lack of effort by the authors to push the article through peer review and update the preprint with the peer-corrections) has led to a cacophony of articles, of highly varying quality. It is a long way from the original LAN-L arXiv server, where quality was driven by the desire to do good science and avoid embarrassing oneself in front of our colleagues. The volume of articles you published, or your H-index (invented in 2005) did not matter, and scientific discovery was less cut-throat and profit-focused. 

But perhaps the most concerning issues today involve the parallel distrust in science and the desire to create ever increasingly complex and larger models, whatever the cost. PhysioNet continues to address these issues. By creating software to accompany the data, we generate examples of truly repeatable science. In addition, through the PhysioNet Challenges, we provide a gold standard for repeatable AI and evaluation of algorithms, as described in section \ref{sec:history}.

The question of how many resources should we throw at a problem is a tricky one. The current zeitgeist aligns with `bigger is better', but this has multiple drawbacks, as we describe in section \ref{section:RecentAI}. 
PhysioNet is, again, continuing to address this, through our publication of highly efficient signal processing algorithms, and the computational limitations we impose during the Challenges. There are also recent trends towards rediscovering many signal processing frameworks (such as the Kalman filter, \cite{xie2021deepfilteringdnncnn}) through machine learning. While it is interesting to imagine we can skip the meticulous hand-crafted design of the old approaches, and just optimized parameters (both in terms of number and value) using ML, there is something to be said for the old ways, including the ability to ensure stability, for example. Perhaps as importantly, in the area of medicine, it is the improved interpretability, much faster execution, the need for less training data, and the ability to deploy them in real time. 

The current incentives in academia and industry tend towards quantity, rather than quality, which is leading to breakdown in trust. The public don't trust our scientific articles, and cherry pick when they do `their own research'. One might argue that open access has helped driven that - democratization of access to data doesn't create more knowledge if you are not trained in how to distinguish the truth from fiction. 
Perhaps, as a community, we need to translate PhysioNet tutorials into well-curate general public science tutorials. Just today, while I was finally putting in enough time to finish this article off, I received an odd email while writing this article from `academia.edu', linking me to an AI-generated podcast of one of my papers. It wasn't terrible, but of course, it wanted money for me to share it. The entity that generated it describes itself as {\it a commercial social networking site for academics to upload and share research papers, build professional profiles, and connect with other researchers.} It is not a university or academic institution, despite its {use of the \it .edu}, which it obtained before the restriction on {\it .edu} domains was put in place. This was the first time, in the 35 years I've been using the academic network, that I have come across a non-academic institution using the edu domain. It was exquisite timing for a company to send a promotional email that implied it represented an academic entity, offering to use AI to automatically productize research I posted for free, and then sell it back to me.
Irony aside, it makes me think that this might actually be a good use of AI - LLMs are actually very good at reworking text and tuning it to be well-received at varying educational levels, if we can curate it effectively. Sadly, academia.edu provided no option to improve the product and help me refine it into something I felt really represented my research. The opportunity is out there, though.

%% file: 2_methods.tex
%Of course, there are far more obvious targets for AI for resources like PhysioNet. 
The confluence of high-performance computing, deep learning, and large ECG databases (hundreds of thousands to millions of recordings) has led to a gold rush in the field, with optimistic claims of `cardiologist-level' diagnostics abilities and prescient predictive abilities. There are too many to cite, and singling out any single article would be an injustice, since there is no single work that seems to stand out. Even more recently, the success of large language models (like ChatGPT) has spawned a trend for foundation models across many health data fields.  

%%%%%%%%%%%%%%%%%%%%
%There is a natural gold rush and groups are branding large models as foundation models, so I want to be clear about the definitions here. 

Before discussing this perspective, it's important we define what a foundation model is, at least for this article.
IBM defines them as: \\
 ''{\it Models that are trained on a broad set of unlabeled data that can be used for different tasks, with minimal fine-tuning ... [and] using self-supervised learning and transfer learning, the model can apply information it’s learned in one situation to another.}' \cite{FoundationModelsIBM} 
%[https://research.ibm.com/blog/what-are-foundation-models]

 This might be somewhat too broad a definition, though, and what exactly is meant by unlabeled is unclear. An autoassociative network is designed to reconstruct the input, and is therefore truly 'unlabeled', and a language model that is designed to learn the next word in a sequence is similar. A model trained on a specific task can still generalize to many other tasks when the training data exhibit broad physiological variation. For example, an ECG foundation model trained to predict chronological age may learn representations reflecting diverse cardiovascular characteristics. Although chronological age is an imperfect surrogate for cardiac health, sufficiently large and representative training data spanning wide physiological diversity may enable the model to capture meaningful structure across the signal, allowing that structure to transfer effectively to other tasks with fine-tuning. 

For the purposes of this discussion, I think it's useful to define two classes of large and generalizable models:
  
{\bf Weak foundation model}:  {\it A machine learning or deep learning model trained on diverse data to support its application across a wide range of downstream tasks. The term is often used to refer to large AI models, though foundation models are more precisely defined by their general-purpose adaptability rather than by size alone.} \cite{CMA}.
%From: Competition and Markets Authority (2023). AI Foundation Models: Initial Report. %https://assets.publishing.service.gov.uk/media/65081d3aa41cc300145612c0/Full_report_.pdf

{\bf Strong foundation models}: {\it A ‘general-purpose AI’ or ‘GPAI’ system. These are capable of a range of general tasks (such as text synthesis, image manipulation, and audio generation. Notable examples are OpenAI’s GPT-3 and GPT-4, foundation models that underpin the conversational chat agent ChatGPT}
\cite{Lovelace}.

Here I use `weak' and `strong' not to mean `bad' or `good', but rather to mean `relaxed' or `stringent' definitions, as in weak and strong stationarity.
In general, I don't think any publication has come close to a strong foundation model in cardiology. Even the current array of weak foundation models lacks key additional characteristics to be considered a foundation model. For one, models have not yet been trained on data spanning a sufficiently broad geographic range. Available data come primarily from a handful of datasets in the United States, one or two from Europe and China, and a single dataset from Brazil. It's difficult to imagine that these databases, skewed largely to middle-aged populations on the East Coast of the USA, represent all of humanity. 
In particular, much of South-East Asia and all of Africa are entirely absent from any models published to date. 
%But these definitions also lack precision.

%I think that's an important caveat.  I think it would be interesting for our journal / the focus issue to explore either weak or strong (or perhaps we should call them narrow vs wide) models with this caveat. But you aren't going to see many strong foundation models, I suspect. Most importantly, all the attempts I've seen so far don't show how the foundation model can be used beyond the data / domain on which it was trained. I think every article needs to show how the model can be easily fine tuned on a new domain. Of course, then you have to define what is a new domain. 
%Moving from ECG to PPG - that works for me. From adult to fetal ECG - probably not. From fetal human to mouse - I'm not sure. Probably not. 

Even if we address these issues, we are still a long way from achieving a {\em strong} foundation model in cardiology capable of a wide range of tasks, such as recommending the next test or medication. 

\subsection{Pathways to (Strong) Foundation Models in Cardiology}
A Strong Foundation Model in cardiology would need to move beyond mere pattern recognition to demonstrate Emergent Clinical Reasoning. This subsection briefly details the technical milestones required to transition from the current {\it Weak} models to a {\it Strong} General Purpose AI (GPAI) system for cardiovascular health:

\begin{itemize}
    \item Longitudinal and Multimodal Data Fusion: A strong foundation model must integrate data across vastly different scales, modalities, and timeframes. We need to move beyond combining raw ECG (or ECG images) with clinical text to include X-ray, PET, CT, MRI, and ultrasound images/videos, as well as social determinants of health, clinical context (time of day, physical activity), and even non-medical data like daily emotional expressions and movement patterns.
    \item The limited temporal memory of current foundation models needs to be addressed, shifting from an ability to learn from or analyze static recordings to longitudinal records that capture the ``natural history'' of a heart. This would incorporate patient history (short- and long-term), extending the context window and allowing the model to distinguish between acute changes and chronic baselines.
    \item Biological Universalism through inclusion of the ``Global Majority''. To achieve {\em Strong} foundation model status, it must generalize across the breadth of human physiological variation. This necessitates active incorporation of data from South-East Asia and Africa.
    \item A critical technical milestone for Strong models includes proving that pretraining on massive datasets actually leads to significant improvements over training a task-specific model directly on a target domain, and the size of the database for pretraining needs to be determined. In particular, a strong model should perform {\em zero-shot} or {\em few-shot} diagnostics on rare cardiac conditions it was never explicitly trained to detect, purely by `understanding' (encoding) the underlying physiological structure of the signal. Proof of this `transfer efficiency' requires benchmark standardization, using sequestered test sets from venues such as the PhysioNet Challenges to generate standardized, independent performance reports.
    \item A Strong model must reconcile the `traditional' approaches of hand-crafted signal processing with modern machine learning. This includes adding in engineering domain knowledge, such as spectral context, and infrequent (and therefore undervalued) features such as QRS detection locations.  
\end{itemize}

There has been some progress on the above issues, such as nascent attempts to incorporate clinical data. Some multimodal models (e.g., echo plus ECG) have been developed, but nothing is particularly convincing as a strong foundation model just yet. 
%There have been some attempts to incorporate clinical data, and others have explored multimodal models (e.g., echocardiography combined with ECG), but none yet provide convincing evidence of a strong foundation model. 
%In particular, current approaches have not demonstrated that pretraining on large datasets prior to transfer learning yields meaningful performance improvements over training models directly within the target domain.
In particular, current models generally fail to demonstrate that the pretraining on large datasets before transfer learning leads to significant performance improvements over just training a more suitable model on the target domain
\cite{He_2019_ICCV,raghu2019transfusion,NEURIPS2020_27e9661e,NEURIPS2020_0607f4c7,bommasani2022opportunitiesrisksfoundationmodels,Nguyen2025}. 
In particular, \cite{raghu2019transfusion} found that pretraining medical imaging models on ImageNet does not significantly improve final performance. However, it does speed up the convergence (training speed) of the model, though this can also be achieved through other methods.
Even more interestingly, contrary to expectations, the authors showed that deep medical models do not heavily reuse features learned from natural images, and smaller, simpler models performed comparably to larger, complex models that were pre-trained. This suggests that massive, pre-trained architectures are over-parameterized and unnecessary for many medical imaging tasks. As \cite{bommasani2022opportunitiesrisksfoundationmodels} note, `we currently lack a clear understanding of how [foundation models] work, when they fail, and what they are even capable of due to their emergent properties. To tackle these questions, we believe much of the critical research on foundation models will require deep interdisciplinary collaboration commensurate with their fundamentally sociotechnical nature.'

%The general lack of enormous diversity in the training data, and lack of unsupervised pretraining and transfer learning. 

%There are also lots of ethical issues around resource usage, bias, and representation. I can't imagine claiming a foundation model trained on 100,000 Americans. What about the rest of the world - China, India, Africa, Latin America. Sure - they are (under) represented in the US, in some weird filtered manner, but it's largely European, with a huge splash of bias in terms of the demographics of the creators and the targets.

%%%%%%%%%%%%%%%%%%%%
\subsection{Foundation and Empire - the New Colonialism or Democratization of ML}

I've already argued that we lack a large enough diversity in our datasets to make appropriate foundation models.  
%Claims of foundation models in cardiology, physiology, and medical data in general, are largely premature because foundation models must be trained on a broad enough range of data so that they may be applied across a wide range of use cases. To do so requires a broad representation of both individuals and medical conditions. 
We might argue that the very latest datasets \cite{Ribeiro2020,HarvardEmoryECG2023}, which draw on diverse populations in the USA and Brazil, are sufficient, but there are many social and structural factors that mean that Africa, Asia, and Latin America are still vastly under-represented in our databases. 

Perhaps just as importantly, the researchers working with these data are primarily drawn from Europe and North America, reflecting structural asymmetries in the global distribution of AI expertise and influence. Beyond implications for equity in employment and commercialization, this concentration risks introducing blind spots in model design and in the prioritization of diagnostic or therapeutic targets \cite{Benjamin2019}. 

Another key issue we face in developing Foundation models is the barrier to entry resulting from the enormous computing power required to build large models. 
Access to local high-performance clusters or centralized cloud-based infrastructure has become the primary pathway for building such models. Instead of democratizing access to data, as was the intent of the pioneers in this field (such as Mark, Moody and Goldberger), 
this trend risks concentrating innovation in the wealthiest institutions. Moreover, even if access is made free, the access becomes meaningless without the finances to run large-scale computing on the data. (Here I mean free as in `free beer', in addition to `open' or free as in `free speech', to paraphrase Stallman and others at %the team that GNU Free Software Foundation 
\cite{gnu_free_software}.)

\subsection{The Carbon Footprint of Success}

The elephant in the room with foundation models is their carbon footprint. Training GPT-3 generated 552 ton of CO$_2$ (tCO$_e$) and ChatGPT has more than 13 million users per day with 5 questions each, so 65 million queries would generate 30 tCO$_e$ per day or 0.5 gCO$_{2e}$ per query \cite{deVries2023}. 
%Current AI technology is on track to annually consume as much electricity as the entire country of Ireland (29.3 terawatt-hours per year).
GTP4 cost 10 to 100 times as much to train (\$100 million), and although training has become more efficient, the carbon footprint is likely to still be an order of magnitude larger \cite{patterson2021carbonemissionslargeneural}.  

The enormous datasets and rich data has provided something of a perverse incentive to use `inappropriate' ML models. In other words, we see teams shoehorning off-the-shelf methods that were developed for a qualitatively different problem, without doing something innovative with the architecture or cost function, or otherwise trying to adapt the models to the problem. A majority of the PhysioNet Challenge entries now use deep learning toolkits, and adapt imaging models and LLMs, which are consuming data (and energy) to compensate for a lack of signal processing or domain knowledge. 
Asymptotically, these approaches may outperform signal processing, but at a significant cost and with a significant potential to identify false (and dangerous) features \footnote{Typical examples of such features are energy peaks in non-physiological spectral regions that are present in an external database used to augment a particular class, but which isn't always present. Such artifacts can arise from differences in acquisition hardware or filtering pipelines. 
A programmer unfamiliar with ECG hardware, and the typical spectral bandwidth, might na\:ively assume these datasets are sufficiently comparable. In reality, all hardware and recording systems have different spectral signatures. Something as simple as a 50 Hz (instead of 60 Hz) mains spike, or a strong baseline wander in a particular database, can become an erroneously triggering feature. While this can be addressed in the judicious choice of a rich enough training and test dataset, the programmer may not have the insight or luxury to do this appropriately.}

We cannot close Pandora's box, and perhaps we don't want to, but we do need to be more thoughtful about how and when we train large models on our vast quantities of data, how and when the models are used, and how data are collected. The developments in distributed learning and edge computing hold enormous promise for energy efficiency, privacy preservation, and the empowerment of healthcare workers in the most remote regions, where AI has the potential to have the most impact.  

\subsection{A Quick Aside on Interpretability}
There's an argument against complex and deep models - that we'll never understand how the classification or predictions are made. I think this is probably overblown. 
I am not sure if there's a single person who can explain how any commercial cardiology algorithm works. Since the algorithms have been modified countless times by so many individuals over so many years, no one person has a complete understanding. The only code I can really explain is the QRS detector and neural network I wrote from scratch in C during my PhD. But that was just me working on that, and I had to think about every operation and memory allocation. I spent time to understand every library I used from `Numerical Recipes in C', poring over it line by line.  After that, I mostly wrote in higher-level languages using many built-in functions. While such languages allow researchers to rapidly develop solutions, it has also led to a lot of sloppiness in the use of inappropriate default settings in functions that people don't take the time to understand. Recently, I rewatched the original Jurassic Park with my children, and this reminded me of the lines spoken by Jeff Goldblum as the `chaotician', explaining the hubris of bringing back dinosaurs \cite{Crichton1990}: 

{\it You know what's wrong with scientific power? It’s a form of inherited wealth. ... Most kinds of power require a substantial sacrifice by whoever wants the power. There is an apprenticeship, a discipline lasting many years. ... you have to put in the time, the practice, the effort. You must give up a lot to get it. It has to be very important to you. And once you have attained it, it’s your power. It can't be given away: it resides in you. It is literally the result of your discipline. Now what is interesting about this process is that, by the time someone has acquired the ability to kill with his bare hands, he has also matured to the point where he won't use it unwisely. So that kind of power has a built-in control. ... But scientific power is like inherited wealth: attained without discipline. You read what others have done, and you take the next step. You can do it very young. You can make progress very fast. There is no discipline lasting many decades. There is no mastery: old scientists are ignored. There is no humility before nature. There is only a get-rich-quick, make-a-name-for-yourself-fast philosophy. ... And because you can stand on the shoulders of giants, you can accomplish something quickly. You don't even know exactly what you have done, but already you have reported it; patented it, and sold it.}

It's easy to dismiss the older generation as Luddites, fearing change. To be clear, I embrace change. I'm looking forward to seeing improvements in vibe coding that accelerate our ability to implement. However, I hope this doesn't come at the cost of sloppy code, and as long as we build in the right unit tests and debugging processes, and continue to teach fundamentals of signal processing, I think we'll be fine. I just hope we don't spoil ourselves with inherited wealth. 

%% file: 5_conclusion.tex
Speculating about the future is always going to lead an author open to ridicule from the inevitable forecasting errors. So I'll confine myself to what I think the future {\it should} look like, and why.

\subsection{Large Models}
It's obvious that the trend for large and foundation models isn't going to stop any time soon, even if the AI bubble bursts. They are just too useful.
It seems likely we'll have a useful generalizable model for some domains (such as the ECG) in just a few years (or even months). But there will be some false starts, and we will need to call them out and force their creators/owners to fix them. Building the right framework to do so will not be easy.

Hopefully, the usual interplay between industry and open academia will enable this, to some extent. Industry is already beginning to rush in, with bold and brash startups, promising a lot (perhaps way too much). This is the roller coaster society has built for itself. Whether we like it or want it, it is hard to avoid. I, for one, am up for the ride. I fell in love with the promise of neural networks in the mid-1990s, and I banged my head on small databases, slow processors, and tiny memory footprints,  learning to develop intuitive feature engineering approaches and kernel tricks while embracing the beauty of electrical engineering. I could taste the promise that was 25 years away.

However, we're still not `there,' and we have an imperative to collect more diverse datasets. Significant holes in public data remain, failing to reflect the diversity of humanity, particularly with sparse sampling of Asian and Latin American populations, and virtually no representation from Africa. The Global Majority rarely appears in our algorithms. That may be politically out of favor in some places these days, but it doesn't negate the scientific imperative that foundation models should sample the full breadth of humans.

Just as importantly, we need to build more diverse teams and promote Global Majority researchers, to prevent group-think. Moreover, governments need to ensure that we don't enact a new age of colonialism. That is, we need to ensure that the Global South controls its data for commercialization, so that its populations directly and substantially benefit from the potential profits of its own data. 

Finally, we need to be more thoughtful about how and when we train large models on our vast quantities of data, how and when the models are used, and how data are collected. The developments in distributed learning and edge computing hold enormous promise for energy efficiency, privacy preservation, and the empowerment of healthcare workers in the most remote regions, where AI has the potential to have the most impact.  

I'll end this section with a counterpoint. Multimodal databases enable multimodal foundation models, which are becoming more and more powerful. We will rapidly move beyond text plus ECG, to add ultrasound, and other imaging, social determinants of health, and many other medical or metadata. Eventually, this will include non-medical data such as emotional expressions during daily living, location, movement, and social interactions. These will take enormous efforts, and the costs may turn out to be prohibitive in my lifetime, or at least until the next unpredictable revolution in algorithmic efficiency, of course\footnote{I'm not optimistic about General-Purpose Machines Quantum Computing by 2035 or 2040, as is often optimistically predicted \cite{groenland2021introduction}. While I readily admit that I haven't done much Quantum Physics since the mid 90's, I do still try to keep up to date, and I'm a big believer in the Kahneman and Tversky's 
notion of the Planning Fallacy: the cognitive bias in which we tend to underestimate the amount of time, costs, and risks required to complete future tasks, even when we know that past similar tasks have taken longer than planned. At the same time, I'm a big believer in Amara's Law \cite{lin2024amaraslaw}
in which we tend to overestimate the effect of a technology in the short run and underestimate its effect in the long run. i.e., we have a short-term over-optimism about AI and our capacity to invent new ways to overcome the limitations (power, capacity, cost), but we are perhaps underestimating the long-term transformative potential of the technology. Here I don't mean AGI or the Terminator, but rather something we cannot yet imagine in the more distant future.}. 

\subsection{PhysioNet and the Public Good}
However, there is an emerging danger to public resources, on which much of the innovation in our field has relied. As government funding is stripped away, the question of how to maintain access to data emerges. It costs substantial time, effort, and money to collect, upload, host, serve (download), and curate the data. Historically, PhysioNet has both acquired and generated data, but the cost to generate large volumes of data is increasing, and the effort to curate complex and large clinical or biological data is prohibitive. 

Effective data curation requires supporting meaningful use rather than merely providing access. Historically, this has been achieved through the development of tutorials, novel signal analysis methods, and open-source software that enable the extraction of information from the datasets (Aim 2 of the PhysioNet grant). It has also been advanced through organized Challenges (Aim 3), which can expose important limitations of the data through both internal validation and community-wide evaluation, effectively crowdsourcing technical scrutiny of the resource. 

But this is not enough these days. The internet is awash with code and data, and it's hard to tell the wood from the trees. 
The research landscape has become an incoherent mass of preprints, GitHub repositories, and poorly curated data that fail to pass the bar for replicable science. 
Even famous peer-reviewed journal articles have succumbed to the trap of profit over quality. When publishers' profit margins approach 40\%, amounting to billions of dollars every year, incentives to publish high-quality research are replaced by headlines, fashion, and  untempered claims of `human-level diagnosis.' Yet the code (and data, if posted in any meaningful way) rarely reflect the results in the publication, and are untrainable on new data. It's not revolutionary to suggest that open not-for-profit public fora with meaningful scientific moderation and curation can address these issues. After all, I'm just describing how journals used to operate prior to the 1950's. The subsequent explosion of scientific research and government funding led commercial publishers, led by figures like Robert Maxwell of Pergamon Press, to capitalize on the high demand for research and the dissemination of its products, transforming scholarly publishing into a highly lucrative industry.

As I've already noted, the PhysioNet Challenges are unique in many ways, but perhaps most importantly, they require the teams to enter code that is fully retrainable on the public training data. Coupled with the scientific publications and annual meetings to discuss the results, the Challenges set a bar for repeatable scientific research.  
Unfortunately, each Challenge takes significant effort to run and, with increasing complexity, continues to increase in cost. 
%and costs much more to run than was anticipated in the Resource grant, something our reviewers saw clearly. We have plugged gaps in resources with generous donations, but donations are not a long-term sustainability plan. 

\subsection{Sustainable Funding Models for Open Access Science} 

There's always the question of how we can sustain resources like PhysioNet.  Either we must seek government funding on a continual basis, or turn the Resource into a fee-for-service entity. Both paths are difficult, and the latter requires a complex and ethically fraught approach to selling access to data, which is, in some sense, antithetical to the whole ethos on which the Resource was established. The Challenges offer a potential pathway forward, since they have been designed to sequester test data for truly independent evaluation of each algorithm. We have long proposed that the Challenges' data could serve as benchmark test sets run independently to generate standardized performance reports suitable for submission to regulatory bodies. However, we also recognize that implementing such an approach would likely require policy support or regulatory incentives, since commercial entities are likely to favor reporting performance based on internally curated data. 

The NIH has continuously funded PhysioNet for almost 30 years, and we hope to continue that for many more. However, there are many offerings, although often far less coherent and focused, that offer access to code, data, and competitions, and the burden of curating data is rising so fast that it's almost impossible to do so properly. There are several approaches to addressing this. One possible model is found in Wikipedia, where a team of vetted community members curate and review the public information.\footnote{As an aside, it's interesting to note that in 1999, the year PhysioNet started, Richard Stallman called for the development of a free online encyclopedia through the means of inviting the public to contribute articles. The resulting {\it GNUPedia} was eventually retired in favour of the emerging {\it Wikipedia}, which had similar aims and was enjoying greater success.} 

PhysioNet has long benefited from contributions from external research groups that bring domain-specific insights to their data. While such collaboration was once relatively uncommon, policies requiring data access upon publication and growing recognition of its importance for reproducibility have accelerated participation. Nevertheless, simply releasing static archives of data and code provides limited long-term value. A more effective model would support living datasets, i.e., resources that can be annotated, discussed, and refined by credentialed members of the research community. Rather than positioning any single group as the definitive authority, such a system would distribute stewardship while retaining editorial oversight to ensure quality and clarity. Editorial responsibility could be shared with the originating data authors and passed on to future domain experts.

Elements of this model already exist within the PhysioNet Challenges through community discussion forums and beta-testing phases, but the approach could be extended substantially with appropriate resources. Challenges could evolve into dynamic tutorial ecosystems that capture collective methodological insight. Emerging tools such as large language models may further help guide users toward relevant prior discussions and solutions, particularly given the frequency with which similar questions recur. 

Another possible approach is to use AI to help enhance the curation of the data, providing a first-pass analysis of common issues. 

\subsubsection{Economic Sustainability, Democratization and Equity.}
It's an open question of how large and diverse enough databases need to be to effectively train world or foundation models. Moreover, if we are to create truly generalizable models with equitable performance across humanity, then the data from the full spectrum of humanity needs to be represented in the databases. Those who need to be represented the most (the global majority) are the least likely to have the resources to collect the data and be represented. When they are, it's often an outsider who funds and collects the data, asserting an implicit (or sometimes explicit) control over the use of the data. I believe we should be working with researchers in lower-resource areas to help secure funding that pays them to collect the data, manage it, and then choose how it is governed. By charging researchers from high-income regions (or organizations) for access, this can subsidize access for those who cannot afford computing or storage and help to support a research and industrial innovation ecosystem around the data. In this way, LMIC researchers can increase the likelihood that they are represented in foundation models and commercial systems without the associated extreme exploitation. With thoughtful application of licensing and co-development/knowledge transfer strategies, we can avoid a digital version of Henrietta Lacks \cite{Russo01102018}. Ideally, we'd be able to create a way for the owner of the data (the person from whom it was recorded) to 'lease' it to a pre-specified group of companies or users for a slice of the income it generates. The data can remain encrypted, or at least private, in a centralized system. As far as I know, George Church's whole genome sequencing company, Nebula Genomics, was the first to pioneer this business model in the biomedical domain. Of course, this doesn't de-risk the likelihood of a company going bankrupt and having its data sold off as an asset, unless there is a strong legal framework to prevent it. Consequently, the contractual `leasing' of the data needs to be carefully considered, with ownership remaining in the country of origin to prevent the on-selling or sub-leasing of the data without the owners' consent. In order to ensure that LMIC data still ends up in the public commons, older data that has lost its premium value could default into a common repository after an embargo period (e.g., 10 years), perhaps in return for some up-front funding to help collect the data.

\subsection{A Better Method for Allocating Grants?}
\label{sect:BetterResourceAllocation}
% Science and a Proposal to Change How Grants are Awarded}

There has been some attention paid to the bias of awarding funding towards the incumbent, well-funded researchers, and some controversy around whether this gives us value for money \cite{Mariethoz2021ImaginaryCarrot}. Although I've been very well funded, like many of my talented colleagues, I have personally felt like my best ideas have often gone ignored and been treated unfairly at review. As \cite{Nicholson2012Conform} note: ``Too many US authors of the most innovative and influential papers in the life sciences do not receive NIH funding''. 

It's been 15 years since \cite{Ioannidis2011FundPeople} suggested that we fund based on the potential of individuals to innovate, rather than on peer review of ideas. The current system stifles competition (when reviewers become territorial) and decreases innovation and risk (with program officers pressured to fund projects that seem the most likely to succeed with positive preliminary data). Yet the whole point of funding research in academia is to enable highly novel ideas and try out innovative pathways that industry would never risk. 
However, funding an individual (or a team) based on track record may exacerbate the advantages of the seasoned researcher with arguably too large a slice of the pie already. One way to address that issue is to focus the track record on the most recent results. I am proposing that we use the PhysioNet Challenges as a vehicle for awarding grants in the following manner:
\begin{itemize}
    \item We tie the award of a prize for the top entries in a Challenge to a requirement to continue performing research on the Challenge data. 
    \item The topic of the research could be almost anything.
    \item Teams would submit their proposal as a short form on the last day of the Challenge (or before).
    \item The Challenge organizers and a small team of domain experts would rate the top 10 scored entries for the novelty of their method (as described in their submitted research preprint) and the novelty of the proposed research topic. 
    \item The top 3 teams rated teams (from this top 10 scored Challenge entries) would be awarded grants sufficient to fund one PhD student or postdoc for 2 years. 
    \item After two years, the awardees must submit any publications, code, results, and updated data/labels back to the Challenge organizers for posting publicly after one year (to give the teams a chance to submit their results for ongoing funding). 
\end{itemize}

This may seem rather controversial, but funding agencies already allow grants to issue internal seed grants, which are internally reviewed by the research team. My proposal takes this idea and extends it to include an objective metric for assessing eligibility for a grant (a good Challenge score), and excludes collaborators of the research team. This increases independence of thought and reduces consolidation of funding in the already well-funded.

As I've already noted, many public competitions do not yield meaningful outcomes \cite{MaierHein2018RankingsCare}.
I hope the structure I have outlined in this article helps to ensure that the data continues to be used in innovative ways, and is built upon in the immediate aftermath of the Challenge - a critical period when so much is known about the data, and can easily be lost in the decompression that follows the intense finale. 
%This would create a significant enhancement of the original data, and provide meaningful developments. 

%Tutorials have always offered a dynamic insight into 

 %\subsection{Final Thoughts}
 
% Perhaps the most exciting thing to see these days is the reinvention of signal processing in the form of data-driven machine learning. It may be a clich\'e for an old person to bemoan that the new generation are just rediscovering all the things we already know. However, there is a justifiable concern that the lack of understanding of sampling theory, aliasing, and transformations into other domains (such as Fourier) will produce inefficient and even dangerous algorithms. But this is the journey of discovery that every new generation must take, and with each generation, we make a new future. Hopefully the great researchers will connect it with the past, though.    
%As a great 20th Century philosopher once said, 
%`in this bright future you can't forget your past' \cite{Marley} ... and we shouldn't.

%% file: 6_TLDR.tex
I realize this article is too long for today's attention span, so I've summarized it here\footnote{with some help of an LLM - the rest of the document, except where directly quoted and attributed, is all my own words.} as a list of 10 bullet points:

\begin{enumerate}
    \item {\bf Evolution of Medical Data Sharing:}
I trace the journey from early physical sharing of medical data (magnetic tapes, CDs) to the current era of internet-enabled, large-scale open access databases, highlighting the transformative impact on research and clinical practice and the key role of the PhysioNet Resource. 

%\item {\bf PhysioNet’s Pioneering Role:}
%I detail how PhysioNet is a central resource in biomedical informatics, especially in cardiology, providing open access to large, well-annotated physiological datasets, software, and tutorials, and catalyzing research through public competitions. 

\item {\bf Growth of Open ECG Databases:}
There has been a superexponential increase in the size and diversity of open-access ECG databases, with PhysioNet hosting over 15 TB of data as of 2025. However, labeling challenges have shifted reliance from expert annotation to algorithmic labeling. 

\item {\bf The PhysioNet Challenges:}
I describe the unique properties of this  annual public competition and how, over the last 25 years, it has driven innovation, dataset creation, and methodological advances. These Challenges emphasize reproducibility, robust evaluation metrics, and community engagement, setting standards for scientific repeatability in AI and data science. 

\item  {\bf Open Science and Publishing Issues:}
I address the publishing industry’s resistance to true openness, profit-driven motives, and the lack of reproducibility in research. PhysioNet’s approach—pairing data with code and tutorials—serves as a model for transparent, repeatable science. 

\item {\bf Lack of Representation from Africa and South East Asia:}
I identify geographic and demographic gaps in current open access medical databases, noting a lack of representation from Africa and South East Asia, pediatric populations, and `normal' populations that reflect the prevelances we encounter in the real world. 

\item {\bf Rise of Foundation Models in Medicine:}
I address the emergence of large AI models (a.k.a. foundation models) in healthcare, noting both their promise and limitations. Notably, current models lack sufficient geographic and demographic diversity, risking bias and limited generalizability. 

\item {\bf Challenges of AI Democratization and the Carbon Footprint:}
I discuss the potential for `new colonialism' in AI, where data and innovation are concentrated in wealthy regions and corporations. I also highlight the significant environmental costs of training large models, advocating for more efficient, distributed approaches like Tiny-ML and edge computing. 

\item {\bf Global Equity and Ethical Considerations:}
I place a strong emphasis on ensuring that data from the Global South is controlled by and benefits the populations from which the data originated, preventing exploitation and promoting equitable access to the fruits of AI and big data in medicine.

\item {\bf Interpretability and Scientific Discipline:}
I discuss the issues of interpretability and scientific rigor, cautioning against over-reliance on built-in functionality of higher-level code without understanding the underlying science, and emphasizing the value of foundational signal processing knowledge. 

\item {\bf Future Directions and Sustainability:}
I  argue for more diverse datasets, inclusive research teams, and sustainable funding models for open resources like PhysioNet. I suggests leveraging the Challenges to create community-driven tutorials and dynamic datasets. I also argue for an innovative grant allocation process via the Challenges to ensure ongoing progress. 
    
\end{enumerate}

If you made it this far, thanks for reading this. I've been trying to finish this for over a year, so it's cathartic to finally write the last words. 

\section*{Acknowledgments} 
PhysioNet began with the core team of Ary Goldberger, Roger Mark, George Moody, Joe Mietus, Madalena Costa, CK Peng, Isaac Henry and Wei Zong. There were many others that contributed at various times, but these are the ones that I remember as key contributors who were there from the start and kept the ship sailing.  
Over the seven years I spent managing the assembly of the MIMIC II database, I was lucky to recruit and supervise Mauro Villarroel, who was key in assembling the relational database (EMR data), and Qiao Li who was key in curating the bedside waveforms. Li-wei Lehman joined our team in the mid-2000's, helping me push much of the MIMIC II output to PhysioNet, including our deidentification software. Of course, there were scores of amazing masters and PhD students that came through the lab. It's easy to find their names from our publications during that period.

I'm particularly grateful to Roger Mark for having taken a chance on me in the early days of PhysioNet, and entrusting me with so many critical things. Roger has been a mentor, colleague, and friend for more than two decades. Likewise, Ary Goldberger has been pivotal with his friendship, support, and personalized limericks. George Moody was one in a billion. His C code was virtually incomprehensible, yet somehow almost supernatural, his affable character a delight, and his deep understanding of the field unsurpassed. Sadly, George passed away in 2021. He was irreplaceable. 
Matt Reyna has been an indispensable co-lead of the Challenges since 2019, taking on much of the day-to-day efforts. It's impossible for me to oversell just how much he contributes. More recently, I've been lucky enough to pull in Reza Sameni, my longtime collaborator of the last two decades. Together with an ever-changing and dedicated team of PhD students, postdocs, and staff engineers, we meet three times a week to plan and execute these Challenges. It's a testament to George that he used to run this almost entirely on his own! Of course, he had the occasional help (I've been contributing for over 20 years myself), and the Challenges have become more complex to support. But still ... George was a miracle worker.

I must thank Madalena Costa, Matt Reyna and Reza Sameni for taking the time to read this article through, and for providing many useful edits and suggestions. 
It's impossible not to mention Pablo Laguna, Leif S\"ornmo, and Lionel Tarassenko - three incredible mentors and giants in their fields that became my good friends. (In fact, it delights me to note that everyone I mentioned above also became a friend who supported me as much as I did them.) Their influence on my research and thinking has been fundamental. 
One of my most important colleagues (and more importantly, beloved friend) on this journey was Patrick McSharry. During my doctoral studies, Lionel paired us up to work on ... well, whatever we wanted. Our first collaboration was our shared love of nonlinear dynamics, which led us to develop an open source ECG simulator, just for fun, and then post it on PhysioNet. If Patrick hadn't suggested posting the code there, I'm not sure I'd be where I am today. Mauro would say the same. We're all sensitive to initial (an ongoing) conditions.
 Of course, there are so many others that have influenced my academic journey (including my brilliant undergraduate physics tutor, Roy Sambles, who set me on this path at the start of my career), and more recently Dave Albert and Joel Xue from Alivecor and GE ... but I have to stop somewhere. 
 
Last, but by no means least, I want to thank all the people and organizations that have been kind enough to put their trust in us and fund our work. 
The work described in this publication was supported in part by the National Institute of Biomedical Imaging and Bioengineering (NIBIB) and by the Director’s Office of Data Science Strategy (ODSS) of the National Institutes of Health (NIH) under Award Number R01EB030362. 
%My work for PhysioNet is funded by the National Institute of Biomedical Imaging and Bioengineering (NIBIB) under NIH grant R01EB030362 (aim three of the three specific aims).
%and the National Center for Advancing Translational Sciences of the National Institutes of Health under Award Number UL1TR002378. 
Funding for the  George B. Moody PhysioNet Challenges has also been provided by Alivecor Inc., Amazon Web Services, Google, the Gordon and Betty Moore Foundation, the IEEE Signal Processing Society, and Mathworks Inc. In addition, the Federation of American Societies for Experimental Biology (FASEB) specifically recognized the Challenges as a recipient of the ``Distinguished Achievement Award for Data Reuse'', as part of the inaugural DataWorks! prize in 2022. The prize money has also been used to support the Challenges. %This award acknowledges the Challenges' long history and success in encouraging the reuse of data to improve health and healthcare.
The content is solely the responsibility of the author and does not necessarily represent the official views of the NIH, my current and past employers, colleagues, or sponsors. I wouldn't be surprised if I've misremembered some things, but in general, I suspect my memory is mostly intact, for now.

%% file: 7_Appendix.tex
%\begin{table*}[h!]
\begin{sidewaystable}
\vspace{0.05in}
\centering
\small  % You can use \footnotesize or \scriptsize for even smaller text
\begin{tabular}{|c|l|c|c|c|c|l|}
\hline
\textbf{Year} & \textbf{Name of} & \textbf{Number of} & \textbf{Sampling} & \textbf{Average} & \textbf{Number} & \textbf{Geographic} \\ 

\textbf{Released} & \textbf{Database} & \textbf{Recordings} & \textbf{Frequency} & \textbf{Length} & \textbf{of Leads} & \textbf{Location} \\ \hline
1980 & MIT-BIH DB  & 48 & 360 Hz & 30 minutes & 2 & NE USA \\ 
 & \cite{Moody1990} &  &  & &  &  \\ \hline
1985 & AHA DB (*) & 80 & 250 Hz & 30 minutes & 2 & Midwest USA \\ 
& \cite{AHA_Database} &  &  & &  &  \\ \hline
1996 & STAFF III DB (**)  & 104 & 1000 Hz & 1-9 minutes & 12 & Appalachia, USA \\  &  \cite{martinez2017staffiii} & & & & & \\ \hline
1997 & QTB DB & 100 & 360 Hz & 15 minutes & 2 & USA + Western EU \\ 
& \cite{laguna1997qt} &  &  & &  &  \\ \hline
2000 & ESC DB & 90 & 250 Hz & 2 hours & 2 & Western Europe \\ 
& \cite{taddei1992european} &  &  & &  &  \\ \hline
2004 & PTB DB & 549 & 1000 Hz & 10-100 seconds & 15 & Germany \\ 
& \cite{bousseljot1995cardiodat} &  &  & &  &  \\ \hline
%2008 & THEW & ~3,000 & 180-1000 Hz & 24 hours & 3-12 & US, Europe, and South America \\ \hline
2018-2020 & CPSC \& CPSC-extra DB & $>$13,000 & 500 Hz & 6-144 seconds & 12 & China \\ 
& \cite{2020ChallengePMEA} &  &  & &  &  \\ \hline
2020 & SaMi-Trop & $\approx$ 2,000 & 400 Hz & 10 seconds & 12 & Brazil \\ 
& \cite{cardoso2016chagas} &  &  & &  &  \\ \hline
2020 & GA ECG DB & $\approx$ 21,000 & 500 Hz & 5-10 seconds & 12 & SE USA \\ 
& \cite{2020ChallengePMEA} &  &  & &  &  \\ \hline
2020 & UMich DB & ~20,000 & 250-500 Hz & 10 seconds & 12 & NE USA \\
& \cite{2020ChallengePMEA} &  &  & &  &  \\ \hline
2020 & CU DB & 45,152 & 500 Hz & 10-100 seconds & 12 & SW US and China \\ 
& \cite{zheng2020optimal} &  &  & &  &  \\ \hline
2020 & CODE 15\% & $\approx$ 200,000 & 400 Hz & 10 seconds & 12 & Brazil \\ 
& \cite{ribeiro2020CODE15} &  &  & &  &  \\ \hline
2022 & PTB-XL DB & 18,869 & 500 Hz & 10-100 seconds & 12 & Germany \\ 
& \cite{wagner2020ptbxl} &  &  & &  &  \\ \hline
2022 & SPH DB & $>$ 24,666  & 200 Hz & 24-hr  & 12 & USA \\
& \cite{liu2022ecgdatabase} &  &  & &  &  \\ \hline
2023 & MIMIC-IV (*) & $~$800,000 & 500 Hz & 10 seconds & 12 & USA \\ 
& \cite{gow2023mimiciv} &  &  & &  &  \\ \hline
2023 & H-E ECG DB & $>$10 million & 500 Hz & 10 seconds & 12 & USA \\ 
& \cite{koscova2024harvardemory} &  &  & &  &  \\ \hline
\end{tabular}
\caption{Key open ECG databases in chronological order. * Not public. ** Not available on PhysioNet until 2017}
\label{table:ecg-databases}
\end{sidewaystable} 
%\end{table*}

\clearpage

\begin{table}[ht]
    \centering
    \caption{Comparison of Annual Public Data Science Competitions\label{tab:DScomps}}
    \renewcommand{\arraystretch}{1.3} % Increases row height for readability
    %\begin{tabular}{p{3cm} p{2cm} p{3cm} p{2cm} p{2cm} p{2cm}}
\begin{tabular}{@{}p{3cm} p{2cm} p{3cm} p{3cm} p{3.5cm}@{}}
        \toprule
          \textbf{Competition / Host} & \textbf{Year Started} & \textbf{Primary \hspace{1cm} Domain} & \textbf{Typical \hspace{1cm} Timeline} & \textbf{Approximate Total Prize} \\
        \midrule
        \textbf{KDD Cup} (ACM SIGKDD) & 1997 & Data Mining / ML & Annually (leading up to the KDD conference in August) & \$30,000 – \$50,000 \\
        \midrule
        \textbf{PhysioNet Challenge} & 1999 & Biomedical Signal Processing & Annual (Jan/Feb through August) & \$3,000 – \$50,000 (Sponsor dependent) \\
        \midrule
        \textbf{Kaggle} (Annuals: March Mania, Santa) & 2010 & General Data Science / ML & Ongoing / Annual (March \& December) & Usually \$25,000 – \$100,000 \\
        \midrule
        \textbf{RecSys Challenge} & 2010 & Recommender Systems & Annual (associated with the ACM RecSys conference in Fall) & \$7,500 – \$10,000 \\
        \midrule
        \textbf{DREAM Challenges} (Sage Bionetworks) & 2011 & Computational Biology & Ongoing / Annual, with phases lasting months & \$0 – \$100,000+ (Often academic prestige) \\
        \midrule
        \textbf{IronViz} (Tableau) & 2011 & Data Visualization & Annual (Qualifiers in late Fall, Final at Tableau Conference) & \$30,000+ (Cash + Charitable Donations) \\
        \midrule
        \textbf{MICCAI} (e.g., BraTS) & 2012 & Medical Imaging / CV & Annual (associated with the MICCAI conference in Fall) & \$0 – \$60,000 (Sponsor dependent) \\
        \bottomrule
    \end{tabular}
\end{table}

\begin{table}[ht]
    \centering
    \caption{Comparison of Annual Public Data Science Competitions (continued)}
    \renewcommand{\arraystretch}{1.3} % Increases row height for readability
    %\begin{tabular}{p{3cm} p{2cm} p{3cm} p{2cm} p{2cm} p{2cm}}
\begin{tabular}{@{}p{3cm} p{2cm} p{3cm} p{3cm} p{3.5cm}@{}}
        \toprule
        \textbf{Competition / Host} & \textbf{Year Started} & \textbf{Primary \hspace{1cm} Domain} & \textbf{Typical \hspace{1cm} Timeline} & \textbf{Approximate Total Prize} \\
        \midrule
        \textbf{DrivenData} & %\textasciitilde
        2014 & Social Good / Public Health & Ongoing / Annual challenges & \$25,000 – \$100,000 (Per Challenge) \\
        \midrule
        \textbf{Capgemini Global DS Challenge} & 2016 & Social Good / Data Science & Annual & \$10,000 – \$20,000 (Cash/Travel Grants) \\
 \midrule
        \textbf{NeurIPS Competition Track} & %\textasciitilde
        2017 & ML / Deep Learning Research & Annual (Conference in December) & \$0 – \$5,000 (Per Track, Academic Focus) \\
        \midrule
        \textbf{IDAO} & 2018 & Data Analysis / Econometrics & Annual (Qualifiers typically start in January) & \$5,000 – \$10,000 (Cash/Scholarships) \\
        \midrule
        \textbf{Machine Hack} & %\textasciitilde 
        2019 & Data Science / ML & Frequent / Annual & \$1,000 – \$15,000 (Varies widely) \\
        \midrule
        \textbf{EY AI \& Data Challenge} & 2021 & Social Good / Sustainability & Annual & \$10,000 – \$20,000 \\
        \bottomrule
    \end{tabular}
\end{table}